\documentclass{tlp}
\usepackage{epsf}
\usepackage{isolatin1}
\usepackage{multicol}
\usepackage{amssymb}

\title[On Applying Or-Parallelism and Tabling to Logic Programs]
      {On Applying Or-Parallelism and Tabling to Logic Programs}

\author[R. Rocha, F. Silva and V. Santos Costa]
       {RICARDO ROCHA, FERNANDO SILVA\\
         DCC-FC \& LIACC\\
         Universidade do Porto, Portugal\\
         \email{\{ricroc,fds\}@ncc.up.pt}
       \and VITOR SANTOS COSTA\\
         COPPE Systems \& LIACC\\
         Universidade do Rio de Janeiro, Brasil\\
         \email{vitor@cos.ufrj.br}
       }

\begin{document}
\maketitle

\begin{abstract}
  Logic Programming languages, such as Prolog, provide a high-level,
  declarative approach to programming. Logic Programming offers great
  potential for implicit parallelism, thus allowing parallel systems
  to often reduce a program's execution time without programmer
  intervention. We believe that for complex applications that take
  several hours, if not days, to return an answer, even limited
  speedups from parallel execution can directly translate to very
  significant productivity gains.
  
  It has been argued that Prolog's evaluation strategy --~SLD
  resolution~-- often limits the potential of the logic programming
  paradigm. The past years have therefore seen widening efforts at
  increasing Prolog's declarativeness and expressiveness. Tabling has
  proved to be a viable technique to efficiently overcome SLD's
  susceptibility to infinite loops and redundant subcomputations.
  
  Our research demonstrates that implicit or-parallelism is a natural
  fit for logic programs with tabling. To substantiate this belief, we
  have designed and implemented an or-parallel tabling engine
  --~OPTYap~-- and we used a shared-memory parallel machine to
  evaluate its performance. To the best of our knowledge, OPTYap is
  the first implementation of a parallel tabling engine for logic
  programming systems. OPTYap builds on Yap's efficient sequential
  Prolog engine.  Its execution model is based on the SLG-WAM for
  tabling, and on the environment copying for or-parallelism.
  
  Preliminary results indicate that the mechanisms proposed to
  parallelize search in the context of SLD resolution can indeed be
  effectively and naturally generalized to parallelize tabled
  computations, and that the resulting systems can achieve good
  performance on shared-memory parallel machines. More importantly, it
  emphasizes our belief that through applying or-parallelism and
  tabling to logic programs the range of applications for Logic
  Programming can be increased.
\end{abstract}


\section{Introduction}

Logic programming provides a high-level, declarative approach to
programming. Arguably, Prolog is the most popular and powerful logic
programming language. Prolog's popularity was sparked by the success
of the sequential execution model presented in 1983 by David H. D.
Warren, the \emph{Warren Abstract Machine}
(\emph{WAM})~\cite{Warren-83}. Throughout its history, Prolog has
demonstrated the potential of logic programming in application areas
such as Artificial Intelligence, Natural Language Processing,
Knowledge Based Systems, Machine Learning, Database Management, or
Expert Systems.

Logic programs are written in a subset of First-Order Logic, Horn
clauses, that has an intuitive interpretation as positive facts and as
rules. Programs use the logic to express the problem, whilst questions
are answered by a resolution procedure with the aid of user
annotations. The combination was summarized by Kowalski's
motto~\cite{Kowalski-79}:
\[algorithm~=~logic~+control\]
Ideally, one would want Prolog programs to be written as logical
statements first, and for control to be tackled as a separate issue.
In practice, the limitations of Prolog's operational semantics, SLD
resolution, mean that Prolog programmers must be concerned with SLD
semantics throughout program development.

Several proposals have been put forth to overcome some of these
limitations and therefore improve the declarativeness and
expressiveness of Prolog. One such proposal that has been gaining in
popularity is \emph{tabling}, also referred to as \emph{tabulation} or
\emph{memoing}~\cite{Michie-68}. In a nutshell, tabling consists of
storing intermediate answers for subgoals so that they can be reused
when a repeated subgoal appears during the resolution process. It can
be shown that tabling based execution models, such as SLG
resolution~\cite{Chen-96}, are able to reduce the search space, avoid
looping, and that they have better termination properties than SLD
based models. For instance, SLG resolution is guaranteed to terminate
for all logical programs with the \emph{bounded term-size
property}~\cite{Chen-96}.

Work on SLG resolution, as implemented in the XSB logic programming
system~\cite{xsb}, proved the viability of tabling technology for
applications such as Natural Language Processing, Knowledge Based
Systems and Data Cleaning, Model Checking, and Program Analysis. SLG
resolution also includes several extensions to Prolog, namely support
for negation~\cite{Apt-94}, hence allowing for novel applications in
the areas of Non-Monotonic Reasoning and Deductive Databases.

One of the major advantages of logic programming is that it is well
suited for parallel execution. The interest in the parallel execution
of logic programs mainly arose from the fact that parallelism can be
exploited \emph{implicitly} from logic programs. This means that
parallelism can be automatically exploited, that is, without input
from the programmer to express or manage parallelism, ideally making
parallel logic programming as easy as logic programming.

Logic programming offers two major forms of implicit parallelism,
\emph{Or-Parallelism} and \emph{And-Parallelism}. Or-parallelism
results from the parallel execution of alternative clauses for a given
predicate goal, while and-parallelism stems from the parallel
evaluation of subgoals in an alternative clause. Some of the most
well-known systems that successfully supported these forms of
parallelism are: Aurora~\cite{Aurora-88} and Muse~\cite{Ali-90a} for
or-parallelism; \&-Prolog~\cite{Hermenegildo-91},
DASWAM~\cite{Shen-92}, and ACE~\cite{Pontelli-97} for and-parallelism;
and Andorra-I~\cite{Costa-91} for or-parallelism together with
and-parallelism. A detailed presentation of such systems and the
challenges and problems in their implementation can be found
in~\cite{Gupta-01}. Arguably, or-parallel systems have been the most
successful parallel logic programming systems so far.  Experience has
shown that or-parallel systems can obtain very good speedups for
applications that require search.  Examples can be found in
application areas such Parsing, Optimization, Structured Database
Querying, Expert Systems and Knowledge Discovery applications.

The good results obtained with parallelism and with tabling rises the
question of whether further efficiency improvements may be achievable
through parallelism. Freire and colleagues were the first to research
this area~\cite{Freire-95}. Although tabling works for both
deterministic and non-deterministic applications, Freire focused on
the search process, because tabling has frequently been used to reduce
the search space. In their model, each tabled subgoal is computed
independently in a separate computational thread, a \emph{generator
thread}. Each generator thread is the sole responsible for fully
exploiting its subgoal and obtain the complete set of answers.
Arguably, Freire's model will work particularly well if we have many
non-deterministic generators. On the other hand, it will not exploit
parallelism if there is a single generator and many non-tabled
subgoals. It also does not exploit parallelism between a generator's
clauses. As we discuss in Section~\ref{section_performance_analysis},
experience has shown that interesting applications do indeed have a
limited number of generators.

Ideally, we would like to exploit maximum parallelism and take maximum
advantage of current technology for tabling and parallel systems. To
exploit maximum parallelism, we would like to exploit parallelism from
both tabled and non-tabled subgoals. Further, we would like to reuse
existing technology for tabling and parallelism. As such, we would
like to exploit parallelism from tabled and non-tabled subgoals
\emph{in much the same way}. As with Freire, we would focus on
or-parallelism first, and we will focus throughout on shared-memory
platforms.

Towards this goal, we proposed two new computational
models~\cite{Rocha-99a}, \emph{Or-Parallelism within Tabling}
(\emph{OPT}) and \emph{Tabling within Or-Parallelism} (\emph{TOP}).
Both models are based on the idea that all open alternatives in the
search tree should be amenable to parallel exploitation, be they from
tabled or non-tabled subgoals. The OPT model further assumes tabling
as the base component of the parallel system, that is, each
\emph{worker}\footnote{The term \emph{worker} is widely used in the
literature to refer to each computational unit contributing to the
parallel execution.} is a full sequential tabling engine. OPT
triggers or-parallelism when workers run out of alternatives to
exploit: at this point, a worker will share part of its SLG
derivations with the other. In contrast, the TOP model represents the
whole SLG forest as a shared search tree, thus unifying parallelism
with tabling. Workers are logically positioned at branches in this
tree. When a branch completes or suspends, workers move to nodes with
open alternatives, that is, alternatives with either open
clauses or new answers stored in the table.

The main contribution of this work is the design and performance
evaluation of what to the best of our knowledge is the first parallel
tabling logic programming system, OPTYap~\cite{Rocha-01}. We chose the
OPT model for two main advantages, both stemming from the fact that
OPT encapsulates or-parallelism within tabling. First, implementation
of the OPT models follows naturally from two well-understood
implementation issues: we need to implement a tabling engine, and then
we need to support or-parallelism. Second, in the OPT model a worker
can keep its nodes \emph{private} until reaching a sharing point. This
is a key issue in reducing parallel overheads. We remark that it is
common in or-parallel works to say that work is initially
\emph{private}, and that is made \emph{public} after sharing.

OPTYap builds on the YapOr~\cite{Rocha-99b} and YapTab~\cite{Rocha-00}
engines. YapOr was previous work on supporting or-parallelism over
Yap's Prolog system~\cite{Costa-99b}. YapOr is based on the
environment copying model for shared-memory machines, as originally
implemented in Muse~\cite{Ali-90b}. YapTab is a sequential tabling
engine that extends Yap's execution model to support tabled evaluation
for definite programs. YapTab's implementation is largely based on the
ground-breaking design of the XSB system~\cite{Sagonas-94,Rao-97},
which implements the SLG-WAM~\cite{Swift-94b,Sagonas-96,Sagonas-98}.
YapTab has been designed from scratch and its development was done
taking into account the major purpose of further integration to
achieve an efficient parallel tabling computational model, whilst
comparing favorably with current \emph{state of the art} technology.
In other words, we aim at respecting the \emph{no-slowdown
principle}~\cite{Hermenegildo-Phd}: our or-parallel tabling system
should, when executed with a single worker, run as fast or faster
than the current available sequential tabling systems. Otherwise,
parallel performance results would not be significant and fair.

In order to validate our design we studied in detail the performance
of OPTYap in shared-memory machines up to 32 workers. The results we
gathered show that OPTYap does indeed introduce low overheads for
sequential execution and that it compares favorably with current
versions of XSB.  Furthermore, the results show that OPTYap maintains
YapOr's speedups for parallel execution of non-tabled programs, and
that there are tabled applications that can achieve very high
performance through parallelism. This substantiates our belief that
tabling and parallelism can together contribute to increasing the
range of applications for Logic Programming.

\section{Tabling for Logic Programs}

The basic idea behind tabling is straightforward: programs are
evaluated by storing newly found answers of current subgoals in an
appropriate data space, called the \emph{table space}. The method then
uses this table to verify whether calls to subgoals are repeated.
Whenever such a repeated call is found, the subgoal's answers are
recalled from the table instead of being re-evaluated against the
program clauses. In practice, two major issues have to be addressed:

\begin{enumerate}
\item What is a repeated subgoal? We may say that a subgoal repeats if
  it is the same as a previous subgoal, up to variable renaming;
  alternatively, we may say it is repeated if it is an instance of a
  previous subgoal. The former approach is known as
  \emph{variant-based tabling}~\cite{Ramakrishnan-99}, the latter as
  \emph{subsumption-based tabling}~\cite{Rao-96}. Variant-based
  tabling has been researched first and is arguably better understood,
  although there has been significant recent progress in
  subsumption-based tabling~\cite{Johnson-99}. We shall use
  variant-based tabling approach in this work.
\item How to execute subgoals? Clearly, we must change the selection
  function and search rule to accommodate for repeated subgoals. In
  particular, we must address the situation where we recursively call
  a tabled subgoal before we have fully tabled all its
  answers. Several strategies to do so have been
  proposed~\cite{Tamaki-86,Vieille-89,Chen-96}. We use the popular SLG
  resolution~\cite{Chen-96} in this work, mainly because this approach
  has good termination properties.
\end{enumerate}

In the following, we illustrate the main principles of tabled
evaluation using SLG resolution through an example.

\subsection{Tabled Evaluation}

\begin{figure}[!ht]
\centerline{
\epsfxsize=12cm
\epsffile{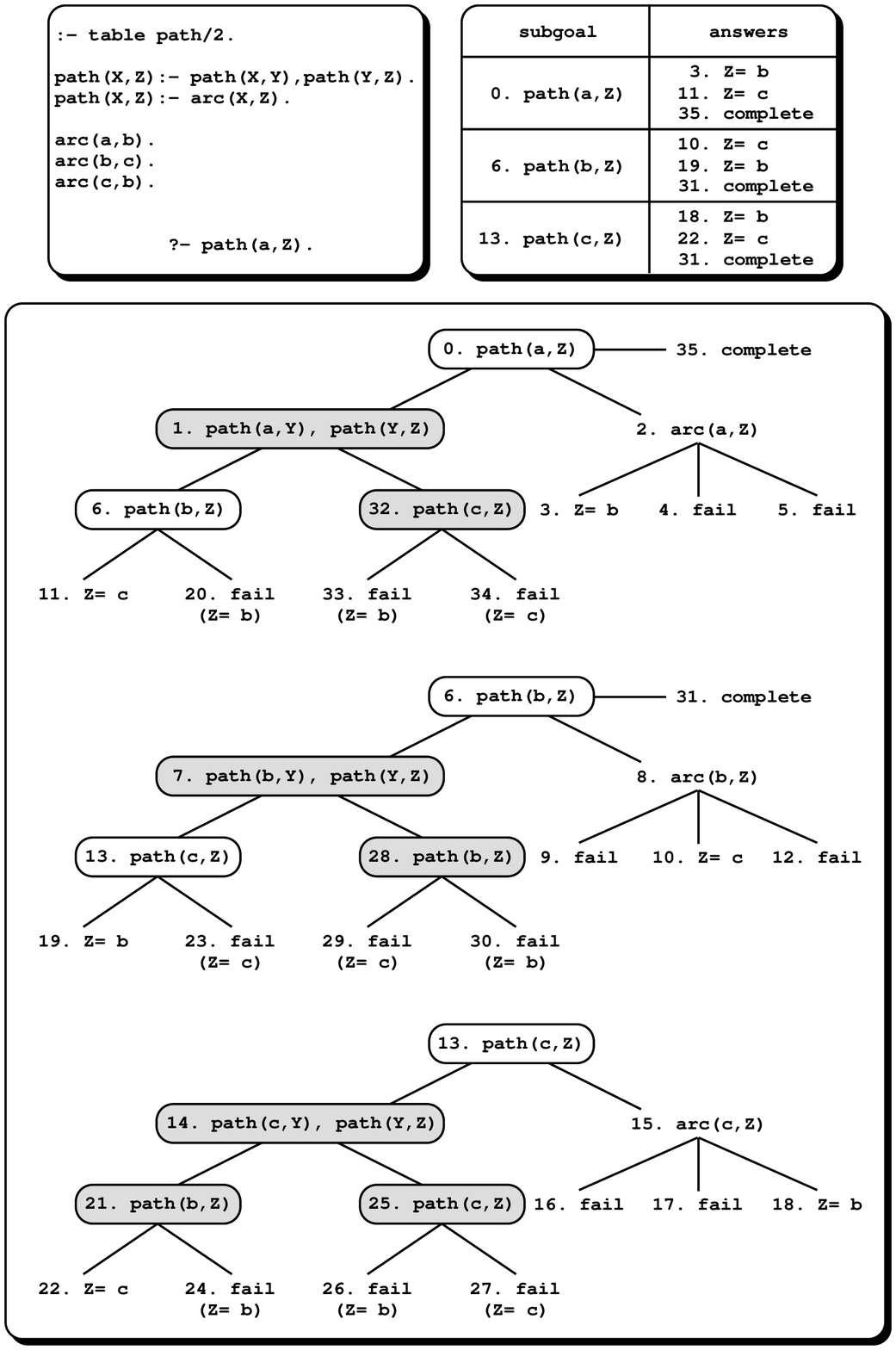}
}
\caption{A finite tabled evaluation.}
\label{fig_finite_SLG_tree}
\end{figure}

Consider the Prolog program of Figure~\ref{fig_finite_SLG_tree}. The
program defines a small directed graph, represented by the
\texttt{arc/2} predicate, with a relation of reachability, given by
the \texttt{path/2} predicate. In this example we ask the query goal
\texttt{?- path(a,Z)} on this program. Note that traditional Prolog
would immediately enter an infinite loop because the first clause of
\texttt{path/2} leads to a repeated call to \texttt{path(a,Z)}. In
contrast, if tabling is applied then termination is ensured. The
declaration \texttt{:- table path/2} in the program code indicates
that predicate \texttt{path/2} should be tabled.
Figure~\ref{fig_finite_SLG_tree} illustrates the evaluation sequence
when using tabling.

At the top, the figure illustrates the program code and the state of
the table space at the end of the evaluation. The main sub-figure
shows the forest of SLG trees for the original query. The topmost tree
represents the original invocation of the tabled subgoal
\texttt{path(a,Z)}. It thus computes all nodes reachable from node
\texttt{a}. As we shall see, computing all nodes reachable from
\texttt{a} requires computing all nodes reachable from \texttt{b} and
all nodes reachable from \texttt{c}. The middle tree represents the
SLG tree \texttt{path(b,Z)}, that is, it computes all nodes reachable
from node \texttt{b}. The bottommost tree represents the SLG tree
\texttt{path(c,Z)}.

Next, we describe in detail the evaluation sequence presented in the
figure. For simplicity of presentation, the root nodes of the SLG
trees \texttt{path(b,Z)} and \texttt{path(c,Z)}, nodes $6$ and $13$,
are shown twice. The numbering of nodes denotes the evaluation
sequence.

Whenever a tabled subgoal is first called, a new tree is added to the
forest of trees and a new entry is added to the table space. We name
first calls to tabled subgoals \emph{generator nodes} (nodes depicted
by white oval boxes). In this case, execution starts with a generator
node, node $0$. The evaluation thus begins by creating a new tree
rooted by \texttt{path(a,Z)} and by inserting a new entry in the table
space for it.

The second step is to resolve \texttt{path(a,Z)} against the first
clause for \texttt{path/2}, creating node $1$. Node $1$ is a variant
call to \texttt{path(a,Z)}. We do not resolve the subgoal against the
program at these nodes, instead we consume answers from the table
space. Such nodes are thus called \emph{consumer nodes} (nodes
depicted by gray oval boxes). At this point, the table does not have
answers for this call. The consumer therefore must \emph{suspend},
either by freezing the whole stacks~\cite{Sagonas-98}, or by copying
the stacks to separate storage~\cite{Demoen-00}.

The only possible move after suspending is to backtrack to node $0$.
We then try the second clause to \texttt{path/2}, thus calling
\texttt{arc(a,Z)}. The \texttt{arc/2} predicate is not tabled, hence
it must be resolved against the program, as Prolog would. We name such
nodes \emph{interior nodes}. The first clause for \texttt{arc/2}
immediately succeeds (step $3$). We return back to the context for the
original goal, obtaining an answer for \texttt{path(a,Z)}, and store
the answer \texttt{Z=b} in the table.

We can now choose between two options. We may backtrack and try the
alternative clauses for \texttt{arc/2}. Otherwise, we may suspend the
current execution, and resume node $1$ with the newly found answer. We
decide to continue exploiting the interior node. Both steps $4$ and
$5$ fail, so we backtrack to node $0$. Node $0$ has no more clauses
left to try, so we try to check whether it has \emph{completed}. It
has not, as node $1$ has not consumed all its answers. We therefore
must resume node $1$. The stacks are thus restored to their state at
node $1$, and the answer \texttt{Z=b} is forwarded to this node. The
subgoal succeeds trivially and we call the continuation,
\texttt{path(b,Z}). This is the first call to \texttt{path(b,Z)}, so
we must create a new tree rooted by \texttt{path(b,Z)} (node $6$),
insert a new entry in the table space for it, and proceed with the
evaluation of \texttt{path(b,Z)}, as shown in the middle tree.

Again, \texttt{path(b,Z)} calls itself recursively, and suspends at
node $7$. We now have two consumers, node $1$ and node $7$. The only
answer in the table was already consumed, so we have to backtrack to
node $6$. This leads to generating a new interior node (node $8$) and
consulting the program for clauses to \texttt{arc(b,Z)}. The first
clause fails (step $9$), but the second clause matches (step $10$).
The answer is returned to node $6$ and stored in the table. We next
have three choices: continue forward execution, backtrack to the open
interior node, or resume the consumer node $7$. In the example we
choose to follow a Prolog-like strategy and continue forward
execution. Step $11$ thus returns the binding \texttt{Z=c} to the
subgoal \texttt{path(a,Z)}. We store this answer in
\texttt{path(a,Z)}'s table entry.

This will be the last answer to \texttt{path(a,Z)}, but we can only
prove so after fully exploiting the tree: we still have an open
interior node (node $8$), and two suspended consumers (nodes $1$ and
$7$). We now choose to backtrack to node $8$, and exploit the last
clause for \texttt{arc/2} (step $12$). At this point we fail all the
way back to node $6$. We cannot complete node $6$ yet, as we have an
unfinished consumer below (node $7$). The only answer in the table for
this consumer is \texttt{Z=c}. We use this answer and obtain a first
call to \texttt{path(c,Z)}.

The new generator, node $13$, needs a new table. Again, we try the
first clause and suspend on the recursive call (node $14$). Next, we
backtrack to the second clause. Resolution on \texttt{arc(c,Z)} (node
$15$) fails twice (steps $16$ and $17$), and then generates an answer,
\texttt{Z=b} (step $18$). We return the answer to node $13$, and store
the answer in the table. Again, we choose to continue forward
execution, thus finding a new answer to \texttt{path(b,Z)}, which is
again stored in the table (step $19$). Next, we continue forward
execution (step $20$), and find an answer to \texttt{path(a,Z)},
\texttt{Z=b}. This answer had already been found at step $3$. SLG
resolution does not store duplicate answers in the table. Instead,
repeated answers \emph{fail}. This is how the SLG-WAM avoids
unnecessary computations, and even looping in some cases.

What to do next? We do not have interior nodes to exploit, so we
backtrack to generator node $13$. The generator cannot complete
because it has a consumer below (node $14$). We thus try to complete
by sending answers to consumer node $14$. The first answer,
\texttt{Z=b}, leads to a new consumer for \texttt{path(b,Z)} (node
$21$). The table has two answers for \texttt{path(b,Z)}, so we can
continue the consumer immediately. This gives a new answer
\texttt{Z=c} to \texttt{path(c,Z)}, which is stored in the table
(step $22$). Continuing forward execution results in the answer
\texttt{Z=c} to \texttt{path(b,Z)} (step $23$). This answer repeats
what we found in step $10$, so we must fail at this point.
Backtracking sends us back to consumer node $21$. We then consume the
second answer for \texttt{path(b,Z)}, which generates a repeated
answer, so we fail again (step $24$). We then try consumer node $14$.
It next consumes the second answer, again leading to repeated
subgoals, as shown in steps $25$ to $27$. At this point we fail back
to node $13$, which makes sure that all answers to the consumers below
(nodes $14$, $21$, and $25$) have been tried. Unfortunately, node $13$
cannot complete, because it depends on subgoal \texttt{path(b,Z)}
(node $21$). Completing \texttt{path(c,Z)} earlier is not safe because
we can loose answers. Note that, at this point, new answers can still
be found for subgoal \texttt{path(b,Z)}. If new answers are found,
consumer node $21$ should be resumed with the newly found answers,
which in turn can lead to new answers for subgoal \texttt{path(c,Z)}.
If we complete sooner, we can loose such answers.

Execution thus backtracks and we try the answer left for consumer node
$7$. Steps $28$ to $30$ show that again we only get repeated answers.
We fail and return to node $6$. All nodes in the trees for node $6$
and node $13$ have been exploited. As these trees do not depend on any
other tree, we are sure no more answers are forthcoming, so at last
step $31$ declares the two trees to be complete, and closes the
corresponding table entries.

Next we backtrack to consumer node $1$. We had not tried \texttt{Z=c}
on this node, but exploiting this answer leads to no further answers
(steps $32$ to $34$). The computation has thus fully exploited every
node, and we can complete the remaining table entry (step $35$).

\subsection{SLG-WAM Operations}

The example showed four new main operations: entering a tabled subgoal;
adding a new answer to a generator; exporting an answer from the
table; and trying to complete the tree. In more detail:

\begin{enumerate}
\item The \emph{tabled subgoal call} operation is a call to a tabled
  subgoal. It checks if a subgoal is in the table, and if not, adds a
  new entry for it and allocates a new generator node (nodes $0$, $6$
  and $13$). Otherwise, it allocates a consumer node and starts
  consuming the available answers (nodes $1$, $7$, $14$, $21$, $25$,
  $28$ and $32$).
\item The \emph{new answer} operation returns a new answer to a
  generator. It verifies whether a newly generated answer is already
  in the table, and if not, inserts it (steps $3$, $10$, $11$, $18$,
  $19$ and $22$). Otherwise, it fails (steps $20$, $23$, $24$, $26$,
  $27$, $29$, $30$, $33$, and $34$).
\item The \emph{answer resolution} operation forwards answers from
  the table to a consumer node. It verifies whether newly found
  answers are available for a particular consumer node and, if any,
  consumes the next one. Otherwise, it schedules a possible resolution
  to continue the execution. Answers are consumed in the same order
  they are inserted in the table. The answer resolution operation is
  executed every time the computation reaches a consumer node.
\item The \emph{completion} operation determines whether a tabled
  subgoal is \emph{completely evaluated}. It executes when we
  backtrack to a generator node and all of its clauses have been
  tried. If the subgoal has been completely evaluated, the operation
  closes its table entry and reclaims space (steps $31$ and $35$).
  Otherwise, it schedules a possible resolution to continue the
  execution.
\end{enumerate}

The example also shows that we have some latitude on where and when to
apply these operations. The actual sequence of operations thus depends
on a \emph{scheduling strategy}. We next discuss the main principles
for completion and scheduling strategies in some more detail.

\subsection{Completion}

Completion is needed in order to recover space and to support
negation. We are most interested on space recovery in this work.
Arguably, in this case we could delay completion until the very end of
execution. Unfortunately, doing so would also mean that we could only
recover space for suspended (consumer) subgoals at the very end of the
execution. Instead we shall try to achieve \emph{incremental
completion}~\cite{Chen-95} to detect whether a generator node has been
fully exploited, and if so to recover space for all its consumers.

\begin{figure}[!ht]
\centerline{
\epsfxsize=7cm
\epsffile{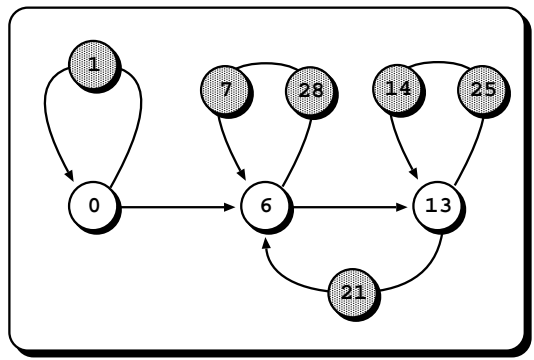}
}
\caption{Node dependencies for the completed graph.}
\label{fig_graph_dependencies}
\end{figure}

Completion is hard because a number of generators may be mutually
dependent. Figure~\ref{fig_graph_dependencies} shows the dependencies
for the completed graph. Node $0$ depends on itself recursively
through consumer node $1$, and on generator node $6$. Node $6$ depends
on itself, consumer nodes $7$ and $28$, and on node $13$. Node $13$
also depends on itself, consumer nodes $14$ and $25$, and on node $6$
through consumer node $21$. There is thus a loop between nodes $6$ and
$13$: if we find a new answer for node $6$, we may get new answers for
node $13$, and so for node $6$.

In general, a set of mutually dependent subgoals forms a
\emph{Strongly Connected Component} (or
\emph{SCC})~\cite{Tarjan-72}. Clearly, we can only complete SCCs
together. We will usually represent an SCC through the oldest
generator. More precisely, the youngest generator node which does not
depend on older generators is called the \emph{leader node}. A leader
node is also the oldest node for its SCC, and defines the current
completion point.
  
XSB uses a stack of generators to detect completion
points~\cite{Sagonas-98}. Each time a new generator is introduced it
becomes the current leader node. Each time a new consumer is
introduced one verifies if it is for an older generator node ${\cal
G}$. If so, ${\cal G}$'s leader node becomes the current leader
node. Unfortunately, this algorithm does not scale well for parallel
execution, which is not easily representable with a single stack.

\subsection{Scheduling}

At several points we had to choose between continuing forward
execution, backtracking to interior nodes, returning answers to
consumer nodes, or performing completion. Ideally, we would like to
run these operations in \emph{parallel}. In a sequential system, the
decision on which operation to perform is crucial to system
performance and is determined by the \emph{scheduling strategy}.
Different scheduling strategies may have a significant impact on
performance, and may lead to different order of answers. YapTab
implements two different scheduling strategies, \emph{batched} and
\emph{local}~\cite{Freire-96}. YapTab's default scheduling strategy is
batched.

Batched scheduling is the strategy we followed in the example: it
favors forward execution first, backtracking to interior nodes next,
and returning answers or completion last. It thus tries to delay the
need to move around the search tree by \emph{batching} the return of
answers. When new answers are found for a particular tabled subgoal,
they are added to the table space and the evaluation continues until
it resolves all program clauses for the subgoal in hand.

Batched scheduling runs all interior nodes before restarting the
consumers. In the worst case, this strategy may result in creating a
complex graph of interdependent consumers. Local scheduling is an
alternative tabling scheduling strategy that tries to evaluate
subgoals as independently as possible, by executing one SCC at a time.
Answers are only returned to the leader's calling environment when its
SCC is completely evaluated.

\section{The Sequential Tabling Engine}

We next give a brief introduction to the implementation of YapTab.
Throughout, we focus on support for the parallel execution of definite
programs.

The YapTab design is WAM based, as is the SLG-WAM. Yap data
structures' are very close to the WAM's~\cite{Warren-83}: there is a
\emph{local stack}, storing both choice points and environment frames;
a \emph{global stack}, storing compound terms and variables; a
\emph{code space area}, storing code and the internal database; a
\emph{trail}; and a \emph{auxiliary stack}. To support the SLG-WAM we
must extend the WAM with a new data area, the \emph{table space}; a
new set of registers, the \emph{freeze registers}; an extension of the
standard trail, the \emph{forward trail}. We must support four new
operations: \emph{tabled subgoal call}, \emph{new answer},
\emph{answer resolution}, and \emph{completion}. Last, we must support
one or several \emph{scheduling strategies}.

We reconsidered decisions in the original SLG-WAM that can be a
potential source of parallel overheads. Namely, we argue that the
stack based completion detection mechanism used in the SLG-WAM is not
suitable to a parallel implementation. The SLG-WAM considers that the
control of leader detection and scheduling of unconsumed answers
should be done at the level of the data structures corresponding to
first calls to tabled subgoals, and it does so by associating
completion frames to generator nodes. On the other hand, YapTab
considers that such control should be performed through the data
structures corresponding to variant calls to tabled subgoals, and thus
it associates a new data structure, the \emph{dependency frame}, to
consumer nodes. We believe that managing dependencies at the level of
the consumer nodes is a more intuitive approach that we can take
advantage of.

The introduction of this new data structure allows us to reduce the
number of extra fields in tabled choice points and to eliminate the
need for a separate completion stack. Furthermore, allocating the data
structure in a separate area simplifies the implementation of
parallelism. We next review the main data structures and algorithms of
the YapTab design. A more detailed description is given
in~\cite{Rocha-PhD}.

\subsection{Table Space}

The table space can be accessed in different ways: to look up if a
subgoal is in the table, and if not insert it; to verify whether a
newly found answer is already in the table, and if not insert it; to
pick up answers to consumer nodes; and to mark subgoals as
completed. Hence, a correct design of the algorithms to access and
manipulate the table data is a critical issue to obtain an efficient
tabling system implementation.

Our implementation of tables uses tries as proposed by Ramakrishnan
\emph{et al.}~\cite{Ramakrishnan-99}. Tries provide complete
discrimination for terms and permit lookup and possibly insertion to
be performed in a single pass through a term. In
section~\ref{section_concurrent_table_access} we discuss how OPTYap
supports concurrent access to tries.

\begin{figure}[!ht]
\centerline{
\epsfxsize=11cm
\epsffile{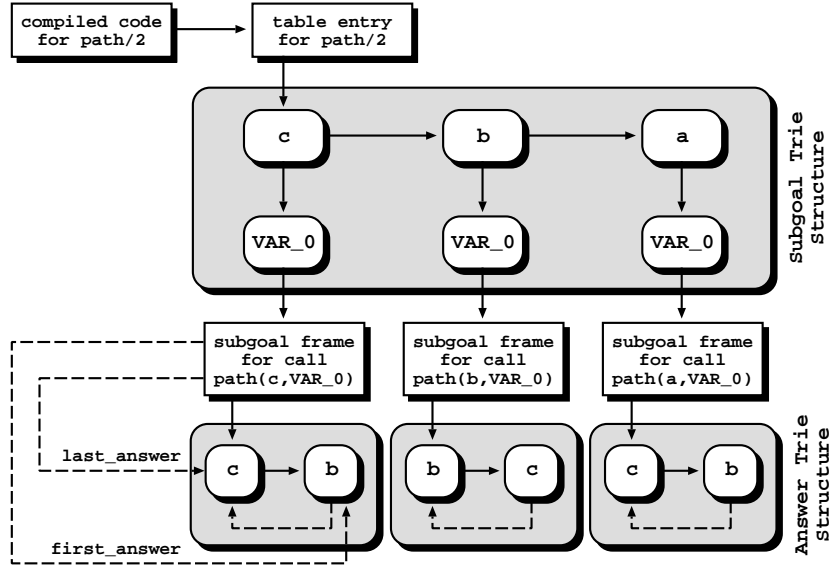}
}
\caption{Using tries to organize the table space.}
\label{fig_tries}
\end{figure}

Figure~\ref{fig_tries} shows the completed table for the query shown
in Figure~\ref{fig_finite_SLG_tree}. Table lookup starts from the
\emph{table entry} data structure. Each table predicate has one such
structure, which is allocated at compilation time. A pointer to the
table entry can thus be included in the compiled code. Calls to the
predicate will always access the table starting from this point.

The table entry points to a tree of trie nodes, the \emph{subgoal trie
  structure}. More precisely, each different call to \texttt{path/2}
corresponds to a unique path through the subgoal trie structure. Such
a path always starts from the table entry, follows a sequence of
subgoal trie data units, the \emph{subgoal trie nodes}, and terminates
at a leaf data structure, the \emph{subgoal frame}.

Each subgoal trie node represents a binding for an argument or
sub-argument of the subgoal. In the example, we have three possible
bindings for the first argument, \texttt{X=c}, \texttt{X=b}, and
\texttt{X=a}. Each binding stores two pointers: one to be followed if
the argument matches the binding, the other to be followed otherwise.

We often have to search through a chain of sibling nodes that
represent alternative paths, e.g., in the query \texttt{path(a,Z)} we
have to search through nodes \texttt{X=c} and \texttt{X=b} until
finding node \texttt{X=a}. By default, this search is done
sequentially. When the chain becomes larger then a threshold value, we
dynamically index the nodes through a hash table to provide direct
node access and therefore optimize the search.

Each subgoal frame stores information about the subgoal, namely an
entry point to its \emph{answer trie structure}. Each unique path
through the answer trie data units, the \emph{answer trie nodes},
corresponds to a different answer to the entry subgoal. All answer
leave nodes are inserted in a linked list: the subgoal trie points at
the first and last entry in this list. Leaves' answer nodes are
chained together in insertion time order, so that we can recover
answers in the same order they were inserted. A consumer node thus
needs only to point at the leaf node for its last consumed answer, and
consumes more answers just by following the chain of leaves.

\subsection{Generator and Consumer Nodes}

\begin{figure}[!ht]
\centerline{
\epsfxsize=12cm
\epsffile{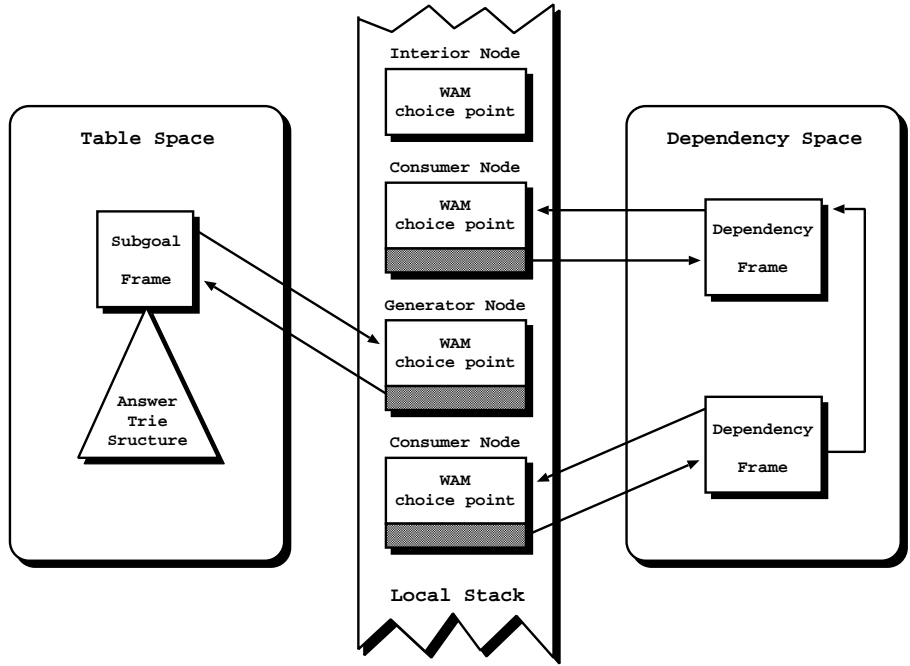}
}
\caption{The nodes and their relationship with the table and dependency spaces.}
\label{fig_nodes_relationships}
\end{figure}

Generator and consumer nodes correspond, respectively, to first and
variant calls to tabled subgoals, while interior nodes correspond to
normal, not tabled, subgoals. Interior nodes are implemented at the
engine level as WAM choice points. To implement generator nodes we
extended the WAM choice points with a pointer to the corresponding
subgoal frame. To implement consumer nodes we use the notion of
\emph{dependency frame}. Dependency frames will be stored in a proper
space, the \emph{dependency space}.
Figure~\ref{fig_nodes_relationships} illustrates how generator and
consumer nodes interact with the table and dependency spaces. As we
shall see in section~\ref{section_leader_nodes}, having a separate
dependency space is quite useful for our copying-based implementation,
although dependency frames could be stored together with the
corresponding choice point in the sequential implementation. All
dependency frames are linked together to form a dependency list of
consumer nodes. Additionally, dependency frames store information
about the last consumed answer for the correspondent consumer node;
and information for detecting completion points, as we discuss next.

\subsection{Leader Nodes}

We need to perform completion in order to recover space and in order
to determine negative loops between subgoals in programs with
negation. In this work we focus on positive programs only, so our goal
will be to recover space. Unfortunately, as an artifact of the
SLG-WAM, it can happen that the stack segments for a SCC ${\cal S}$
remain within the stack segments for another SCC ${\cal S'}$. In such
cases, ${\cal S}$ cannot be recovered in advance when completed, and
thus, recovering its space must be delayed until ${\cal S'}$ also
completes. To approximate SCCs in a stack-based implementation,
Sagonas~\cite{Sagonas-PhD} denotes a set of SCCs whose space must be
recovered together as an \emph{Approximate SCC} or \emph{ASCC}. For
simplicity, in the following we will use the SCC notation to refer to
both ASCCs and SCCs.

The completion operation takes place when we backtrack to a generator
node that \textbf{(i)} has exhausted all its alternatives and that
\textbf{(ii)} is as a leader node (remember that the youngest
generator node which does not depend on older generators is called a
leader node). We designed novel algorithms to quickly determine
whether a generator node is a leader node. The key idea in our
algorithms is that each dependency frame holds a pointer to the
resulting leader node of the SCC that includes the correspondent
consumer node. Using the leader node pointer from the dependency
frames, a generator node can quickly determine whether it is a leader
node. More precisely, in our algorithm, a generator ${\cal L}$ is a
leader node when either \textbf{(a)} ${\cal L}$ is the youngest tabled
node, or \textbf{(b)} the youngest consumer that says ${\cal L}$ is
the leader.

Our algorithm thus requires computing leader node information whenever
creating a new consumer node ${\cal C}$. We proceed as follows. First,
we hypothesize that the leader node is ${\cal C}$'s generator, say
${\cal G}$. Next, for all consumer nodes older than ${\cal C}$ and
younger than ${\cal G}$, we check whether they depend on an older
generator node. Consider that there is at least one such node and that
the oldest of these nodes is ${\cal G'}$. If so then ${\cal G'}$ is
the leader node. Otherwise, our hypothesis was correct and the leader
node is indeed ${\cal G}$. Leader node information is implemented as a
pointer to the choice point of the newly computed leader node.

\begin{figure}[!ht]
\centerline{
\epsfxsize=12cm
\epsffile{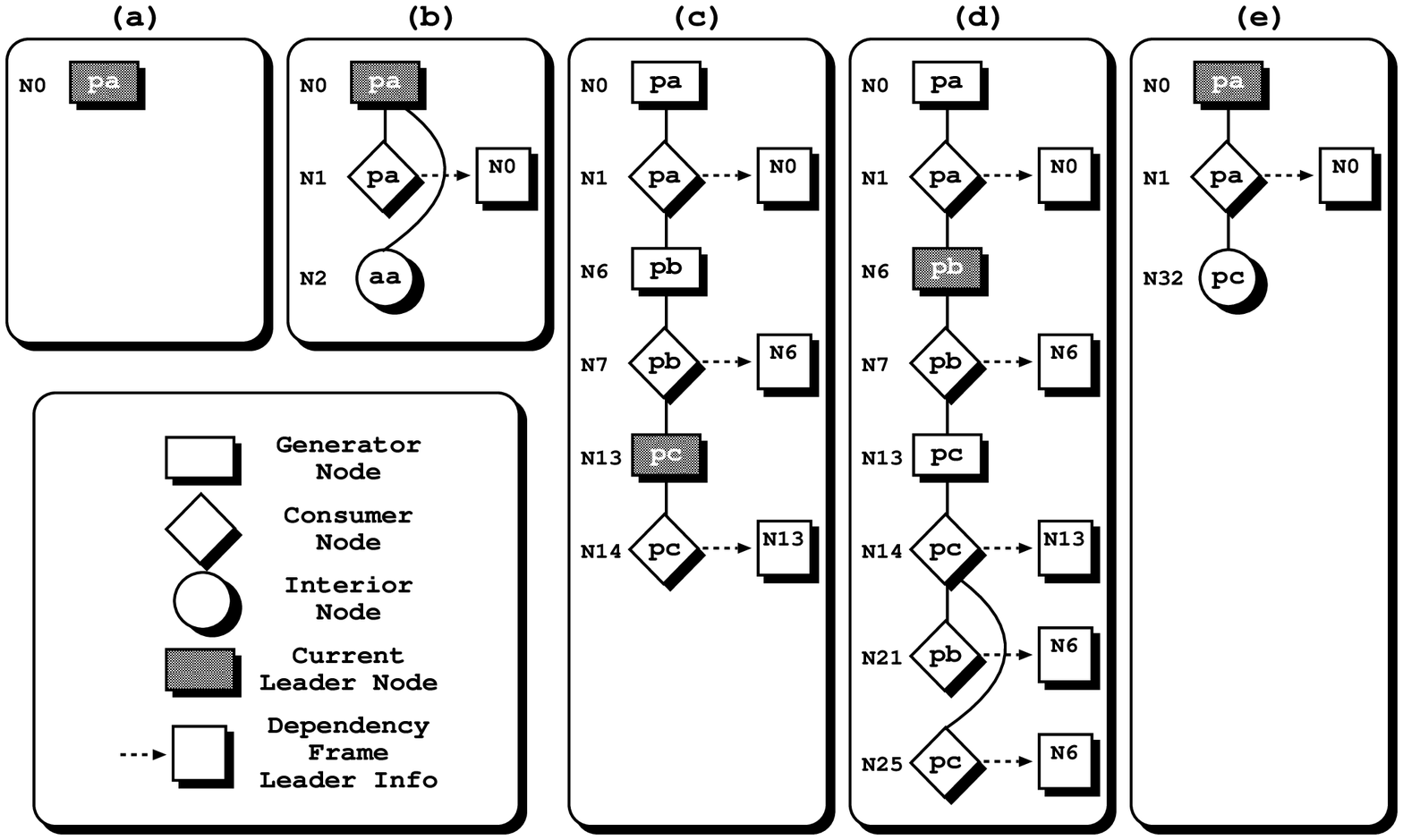}
}
\caption{Spotting the current leader node.}
\label{fig_spotting_current_leader}
\end{figure}

Figure~\ref{fig_spotting_current_leader} uses the example from
Figure~\ref{fig_finite_SLG_tree} to illustrate the leader node
algorithm. For compactness, the figure presents calls to
\texttt{path(a,Z)}, \texttt{path(b,Z)}, \texttt{path(c,Z)} and
\texttt{arc(a,Z)}, as \texttt{pa}, \texttt{pb}, \texttt{pc}, and
\texttt{aa}, respectively. Figure~\ref{fig_spotting_current_leader}(a)
shows the initial configuration. The generator node ${\cal N}_0$ is
the current leader node because it is the only subgoal. Figure
\ref{fig_spotting_current_leader}(b) shows the dependency graph after
creating node ${\cal N}_2$. First, we called a variant of
\texttt{path(a,Z)}, and allocated the corresponding dependency frame.
${\cal N}_0$ is the generator node for the variant call
\texttt{path(a,Z)}, ${\cal N}_0$ is the leader node for ${\cal
  N}_1$'s. ${\cal N}_1$ then suspended, we backtracked to ${\cal N}_0$
and called \texttt{arc(a,Z)}. As \texttt{arc(a,Z)} is not tabled, we
had to allocate an interior node for ${\cal N}_2$.

Figure \ref{fig_spotting_current_leader}(c) shows the graph after we
created node ${\cal N}_{14}$. We have already created first and
variant calls to subgoals \texttt{path(b,Z)} and \texttt{path(c,Z)}.
Two new dependency frames were allocated and initialized. We thus have
three SCCs on stack: one per generator. The youngest SCC on stack is
for subgoal \texttt{path(c,Z)}. As a result, the current leader node
for the new set of nodes becomes ${\cal N}_{13}$. This is the one
referred in the youngest dependency frame.

Figure \ref{fig_spotting_current_leader}(d) shows the interesting case
where tabled nodes exist between a consumer and its generator. In the
example, consumer node ${\cal N}_{21}$, has two consumers, ${\cal
  N}_7$ and ${\cal N}_{14}$, separating it from its generator, ${\cal
  N}_6$.  As both consumers do not depend on nodes older than ${\cal
  N}_6$, the leader node for ${\cal N}_{21}$ is still ${\cal N}_6$,
and ${\cal N}_6$ becomes the current leader node. This situation
represents the point at which subgoal \texttt{path(c,Z)} starts
depending on subgoal \texttt{path(b,Z)} and their SCCs are merged
together. Next, we allocated consumer node ${\cal N}_{25}$. Nodes
${\cal N}_{14}$ and ${\cal N}_{21}$ are between ${\cal N}_{25}$ and
the generator ${\cal N}_{13}$. Our algorithm says that since ${\cal
  N}_{21}$ depends on an older generator node, ${\cal N}_6$, the
leader node information for ${\cal N}_{25}$ is also ${\cal N}_6$. As a
result, ${\cal N}_6$ remains the current leader node.

Finally, Figure \ref{fig_spotting_current_leader}(e) shows the point
after the subgoals \texttt{path(b,Z)} and \texttt{path(c,Z)} have
completed and the segments belonging to their SCC have been
released. The computation switches back to ${\cal N}_1$, consumes the
next answer and calls \texttt{path(c,Z)}. At this point,
\texttt{path(c,Z)} is already completed, and thus we can avoid
consumer node allocation and instead perform what is called the
\emph{completed table optimization}~\cite{Sagonas-98}. This
optimization allocates a node, similar to an interior node, that will
consume the set of found answers executing compiled code directly from
the trie data structure associated with the completed
subgoal~\cite{Ramakrishnan-99}.

\subsection{Completion and Answer Resolution}
\label{section_completion_answer_resolution}

After backtracking to a leader node, we must check whether all younger
consumer nodes have consumed all their answers. To do so, we walk the
chain of dependency frames looking for a frame which has not yet
consumed all the generated answers. If there is such a frame, we
should resume the computation of the corresponding consumer node. We
do this by restoring the stack pointers and backtracking to the node.
Otherwise, we can perform completion. This includes \textbf{(i)}
marking as complete all the subgoals in the SCC; \textbf{(ii)}
deallocating all younger dependency frames; and \textbf{(iii)}
backtracking to the previous node to continue the execution.

Backtracking to a consumer node results in executing the answer
resolution operation. The operation first checks the table space for
unconsumed answers. If there are new answers, it loads the next
available answer and proceeds. Otherwise, it backtracks again. If this
is the first time that backtracking from that consumer node takes
place, then it is performed as usual. Otherwise, we know that the
computation has been resumed from an older generator node ${\cal G}$
during an unsuccessful completion operation. Therefore, backtracking
must be done to the next consumer node that has unconsumed answers and
that is younger than ${\cal G}$. If no such consumer node can be
found, backtracking must be done to the generator node ${\cal G}$.

The process of resuming a consumer node, consuming the available set
of answers, suspending and then resuming another consumer node can be
seen as an iterative process which repeats until a fixpoint is
reached. This fixpoint is reached when the SCC is completely
evaluated.

\section{Or-Parallelism within Tabling}

The first step in our research was to design a model that would allow
concurrent execution of all available alternatives, be they from
generator, consumer or interior nodes. We researched two designs: the
TOP (Tabling within Or Parallelism) model and the OPT (Or-Parallelism
within Tabling) model.

Parallelism in the TOP model is supported by considering that a
parallel evaluation is performed by a set of independent WAM engines,
each managing an unique branch of the search tree at a time. These
engines are extended to include direct support to the basic table
access operations, that allow the insertion of new subgoals and
answers. When exploiting parallelism, some branches may be
\emph{suspended}. Generator and interior nodes suspend alternatives
because we do not have enough processors to exploit them all. Consumer
nodes may also suspend because they are waiting for more answers.
Workers move in the search tree, looking for points where they can
exploit parallelism.

Parallel evaluation in the OPT model is done by a set of independent
tabling engines that \emph{may} share different common branches of the
search tree during execution. Each worker can be considered a
sequential tabling engine that fully implements the tabling
operations: access the table space to insert new subgoals or answers;
allocate data structures for the different types of nodes; suspend
tabled subgoals; resume subcomputations to consume newly found
answers; and complete private (not shared) subgoals. As most of the
computation time is spent in exploiting the search tree involved in a
tabled evaluation, we can say that tabling is the base component of
the system.

The or-parallel component of the system is triggered to allow
synchronized access to the shared parts of the execution tree, in
order to get new work when a worker runs out of alternatives to
exploit, and to perform completion of shared subgoals. Unexploited
alternatives should be made available for parallel execution,
regardless of whether they originate from generator, consumer or
interior nodes. From the viewpoint of SLG resolution, the OPT
computational model generalizes the Warren's multi-sequential engine
framework for the exploitation of or-parallelism. Or-parallelism stems
from having several engines that implement SLG resolution, instead of
implementing Prolog's SLD resolution.

\begin{figure}[!ht]
\centerline{
\epsfxsize=10cm
\epsffile{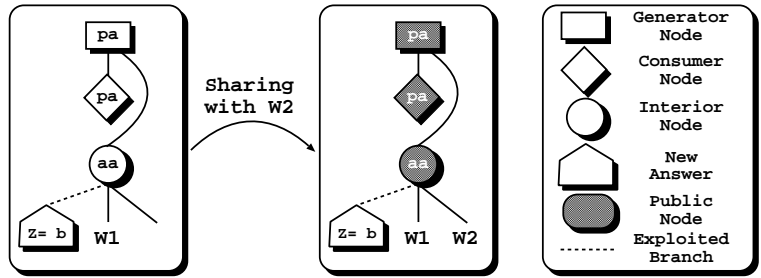}
}
\caption{Exploiting parallelism in the OPT model.}
\label{fig_opt_example}
\end{figure}

We have already seen that the SLG-WAM presents several opportunities
for parallelism. Figure~\ref{fig_opt_example} illustrates how this
parallelism can be specifically exploited in the OPT model. The
example assumes two workers, ${{\cal W}_1}$ and ${{\cal W}_2}$, and
the program code and query goal from Figure~\ref{fig_finite_SLG_tree}.
For simplicity, we use the same abbreviation introduced in
Figure~\ref{fig_spotting_current_leader} to denote the subgoals.

Consider that worker ${\cal W}_1$ starts the evaluation. It first
allocates a generator and a consumer node for tabled subgoal
\texttt{path(a,Z)}. Because there are no available answers for
\texttt{path(a,Z)}, it backtracks. The next alternative leads to a
non-tabled subgoal \texttt{arc(a,Z)} for which we create an interior
node. The first alternative for \texttt{arc(a,Z)} succeeds with the
answer \texttt{Z=b}. The worker inserts the newly found answer in the
table and starts exploiting the next alternative for
\texttt{arc(a,Z)}. This is shown in the left sub-figure. At this
point, worker ${\cal W}_2$ requests for work. Assume that worker
${\cal W}_1$ decides to share all of its private nodes. The two
workers will share three nodes: the generator and consumer nodes for
\texttt{path(a,Z)}, and the interior node for \texttt{arc(a,Z)}.
Worker ${\cal W}_2$ takes the next unexploited alternative of
\texttt{arc(a,Z)} and from now on, either worker can find further
answers for \texttt{path(a,Z)} or resume the shared consumer node.

The OPT model offers two important advantages over the TOP model.
First, OPT reduces to a minimum the overlap between or-parallelism and
tabling. Namely, as the example shows, in OPT it is straightforward to
make nodes public only when we want to share them. This is very
important because execution of private nodes is almost as fast as
sequential execution. Second, OPT enables different data structures
for or-parallelism and for tabling. For instance, one can use the
SLG-WAM for tabling, and environment copying or binding arrays for
or-parallelism.

The question now is whether we can achieve an implementation of the
OPT model, and whether that implementation is \emph{efficient}. We
implemented OPTYap in order to answer this question. In OPTYap,
tabling is implemented by freezing the whole stacks when a consumer
blocks. Or-parallelism is implemented through copying of stacks. More
precisely, we optimize copying by using \emph{incremental copying},
where workers only copy the differences between their stacks. We
adopted this framework because environment copying and the SLG-WAM
are, respectively, two of the most successful or-parallel and tabling
engines. In our case, we already had the experience of implementing
environment copying in the Yap Prolog, the YapOr system, with
excellent performance results~\cite{Rocha-99b}. Adopting YapOr for the
or-parallel component of the combined system was therefore our first
choice.

Regarding the tabling component, an alternative to freezing the stacks
is copying them to a separate storage as in CHAT~\cite{Demoen-00}. We
found two major problems with CHAT. First, to take best advantage of
CHAT we need to have separate environment and choice point stacks, but
Yap has an integrated local stack. Second, and more importantly, we
believe that CHAT is less suitable than the SLG-WAM to an efficient
extension to or-parallelism because of its incremental completion
technique. CHAT implements incremental completion through an
incremental copying mechanism that saves intermediate states of the
execution stacks up to the nearest generator node. This works fine for
sequential tabling, because leader nodes are always generator nodes.
However, as we will see, for parallel tabling this does not hold
because any public node can be a potential leader node. To preserve
incremental completion efficiency in a parallel tabling environment,
incremental saving should be performed up to the parent node, as
potentially it can be a leader node. Obviously, this node-to-node
segmentation of the incremental saving technique will degrade the
efficiency of any parallel system.

\section{The Or-Parallel Tabling Engine}

The OPT model requires changes to both the initial designs for
parallelism and tabling. As we enumerated next, support or-parallelism
plus tabling requires changes to memory allocation, table access, the
completion algorithm. We must further ensure that environment copying
and tabling suspension do not interfere. Or-parallelism issues refer
to scheduling and to speculative work. In more detail:

\begin{enumerate}
\item We must support parallel memory allocation and deallocation of
  the several data structures we use. Fortunately, most of our data
  structures are fixed-sized and parallel memory allocation can be
  implemented efficiently.
\item We must allow for several workers to concurrently read and
  update the table. To do so workers need to be able to lock the
  table. As we shall see finer locking allows for more parallelism,
  but coarser locking has less overheads.
\item OPTYap uses the copying model, where workers do not see the
  whole search tree, but instead only the branches corresponding to
  their current SLG-WAM. It is thus possible that a generator may not
  be in the stacks for a consumer (and vice-versa). We show that one
  can generalize the concept of leader node for such cases, and that
  such a generalization still gives a conservative approximation for a
  SCC. Completion can thus be performed when we are the last worker
  backtracking to the generalized leader nodes, and there is no more
  work below. The first condition can be easily checked through the
  or-parallel machinery. The second condition uses the sequential
  tabling machinery.
\item Or-parallelism and tabling are not strictly orthogonal. More
  precisely, naively sharing or-parallel work might result in
  overwriting suspended stacks. Several approaches may be used to
  tackle this problem, we have proposed and implemented a suspension
  mechanism that gives maximum scheduling flexibility.
\item Scheduling or-parallel work in our system is based on the Muse
  scheduler~\cite{Ali-90b}. Intuitively this corresponds to a form of
  hierarchical scheduling, where we favor tabled scheduling
  operations, and resort to the more expensive or-parallel scheduling
  when no tabling operations are available. Other approaches are
  possible, but this one has served OPTYap well so far. We also
  discuss how moving around the shared parts of the search tree
  changes in the presence of parallelism.
\item Last, we briefly discuss pruning issues. Although pruning in the
  presence of tabling is a complex issue~\cite{Guo-02,Castro-03}, we
  still should execute correctly for non-tabled regions of the search
  tree (interior nodes).
\end{enumerate}

We next discuss these issues in some detail, presenting the general
execution framework.

\subsection{Memory Organization}

\begin{figure}[!ht]
\centerline{
\epsfxsize=7cm
\epsffile{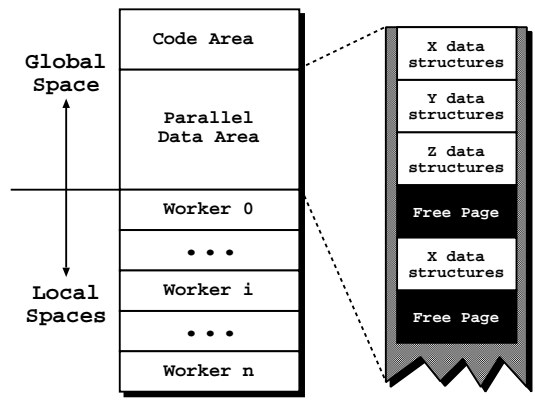}
}
\caption{Memory organization in OPTYap.}
\label{fig_optyap_memory}
\end{figure}

In OPTYap, memory is divided into a \emph{global} addressing space and
a collection of \emph{local} spaces, as illustrated in
Figure~\ref{fig_optyap_memory}. The global space includes the code
area and a parallel data area that consists of all the data structures
required to support concurrent execution. Each local space represents
one system worker and it contains the four WAM execution stacks
inherited from Yap: global stack, local stack, trail, and auxiliary
stack.

The parallel data area includes the table and dependency spaces
inherited from YapTab, and the \emph{or-frame space}~\cite{Ali-90b}
inherited from YapOr to synchronize access to shared nodes.
Additionally, we have an extra data structure to preserve the stacks
of suspended SCCs (further details in section~\ref{section_scc}).
Remember that we use specific extra fields in the choice points to
access the data structures in the parallel data area. When sharing
work, the execution stacks of the sharing worker are copied from its
local space to the local space of the requesting worker. The data
structures from the parallel data area associated with the shared
stacks are automatically inherited by the requesting worker in the
copied choice points.

The efficiency of a parallel system largely depends on how concurrent
handling of shared data is achieved and synchronized. Page faults and
memory cache misses are a major source of overhead regarding data
access or update in parallel systems. OPTYap tries to avoid these
overheads by adopting a page-based organization scheme to split memory
among different data structures, in a way similar to Bonwick's Slab
memory allocator~\cite{Bonwick-94}. Each memory page of the parallel
data area only contains data structures of the same type. Whenever a
new request for a data structure of type ${\cal T}$ appears, the next
available structure on one of the ${\cal T}$ pages is returned. If
there are no available structures in any ${\cal T}$ page, then one of
the free pages is made to be of type ${\cal T}$. A page is freed when
all its data structures are released. A free page can be immediately
reassigned to a different structure type.

\subsection{Concurrent Table Access}
\label{section_concurrent_table_access}

Our experience showed that the table space is the major data area open
to concurrent access operations in a parallel tabling environment. To
maximize parallelism, whilst minimizing overheads, accessing and
updating the table space must be carefully controlled. Reader/writer
locks are the ideal implementation scheme for this purpose. In a
nutshell, we can say that there are two critical issues that determine
the efficiency of a locking scheme for the table. One is the
\emph{lock duration}, that is, the amount of time a data structure is
locked. The other is the \emph{lock grain}, that is, the amount of
data structures that are protected through a single lock request. It
is the balance between lock duration and lock grain that compromises
the efficiency of different table locking approaches. For instance, if
the lock scheme is short duration or fine grained, then inserting many
trie nodes in sequence, corresponding to a long trie path, may result
in a large number of lock requests. On the other hand, if the lock
scheme is long duration or coarse grain, then going through a trie
path without extending or updating its trie structure, may
unnecessarily lock data and prevent possible concurrent access by
others.

Unfortunately, it is impossible beforehand to know which locking
scheme would be optimal. Therefore, in OPTYap we experimented with
four alternative locking schemes to deal with concurrent accesses to
the table space data structures, the \emph{Table Lock at Entry Level}
scheme, TLEL, the \emph{Table Lock at Node Level} scheme, TLNL, the
\emph{Table Lock at Write Level} scheme, TLWL, and the \emph{Table
Lock at Write Level - Allocate Before Check} scheme, TLWL-ABC.

The TLEL scheme essentially allows a single writer per subgoal trie
structure and a single writer per answer trie structure. The main
drawback of TLEL is the contention resulting from long lock
duration. The TLNL enables a single writer per chain of sibling nodes
that represent alternative paths from a common parent node. The TLWL
scheme is similar to TLNL in that it enables a single writer per chain
of sibling nodes that represent alternative paths to a common parent
node. However, in TLWL, the common parent node is only locked when
writing to the table is likely. TLWL also avoids the TLNL memory usage
problem by replacing trie node lock fields with a global array of lock
entries. Last, the TLWL-ABC scheme anticipates the allocation
and initialization of nodes that are likely to be inserted in the
table space before locking.

Through experimentation, we observed that the locking schemes, TLWL
and TLWL-ABC, present the best speedup ratios and they are the only
schemes showing scalability. Since none of these two schemes clearly
outperform the other, we assumed TLWL as the default. The observed
slowdown with higher number of workers for TLEL and TLNL schemes is
mainly due to their locking of the table space even when writing is
not likely. In particular, for repeated answers they pay the cost of
performing locking operations without inserting any new trie node. For
these schemes the number of potential contention points is
proportional to the number of answers found during execution, being
they unique or redundant.

\subsection{Leader Nodes}
\label{section_leader_nodes}

\begin{figure}[!ht]
\centerline{
\epsfxsize=7cm
\epsffile{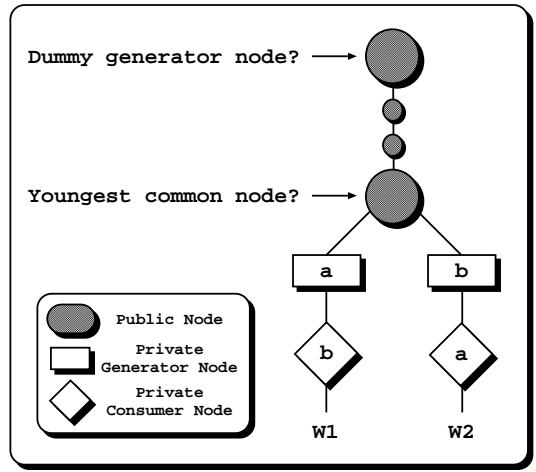}
}
\caption{At which node should we check for completion?}
\label{fig_guess_leader}
\end{figure}

Or-parallel systems execute alternatives early. As a result, different
workers may execute the generator and the consumer subgoals. In fact,
it is possible that generators will execute earlier, and in a
different branch than in sequential execution. As
Figure~\ref{fig_guess_leader} shows, this may induce complex
dependencies between workers, therefore requiring a more elaborate
completion algorithm that may involve branches from several workers.

In this example, worker ${\cal W}_1$ takes the leftmost alternative
while worker ${\cal W}_2$ takes the rightmost from the youngest common
node. While exploiting their alternatives, ${\cal W}_1$ calls a tabled
subgoal \texttt{a} and ${\cal W}_2$ calls a tabled subgoal \texttt{b}.
As this is the first call to both subgoals, a generator node is stored
for each one. Next, each worker calls the tabled subgoal firstly
called by the other, and two consumer nodes, one per worker, are
therefore allocated. At this point both workers hold a consumer node
while not having the corresponding generator node in their branches.
Conversely, the owner of each generator node has consumer nodes being
executed by a different worker. The question is where should we check
for completion? Intuitively, we would like to choose a node that is
common to both branches and the youngest common node seems the better
choice. But that node is not a generator node!

We could avoid this problem by disallowing consumer nodes for
generator nodes on other branches. Unfortunately, such a solution
would severely restrict parallelism. Our solution was therefore to
\emph{allow completion at all kind of public nodes}.

\begin{figure}[!ht]
\centerline{
\epsfxsize=11cm
\epsffile{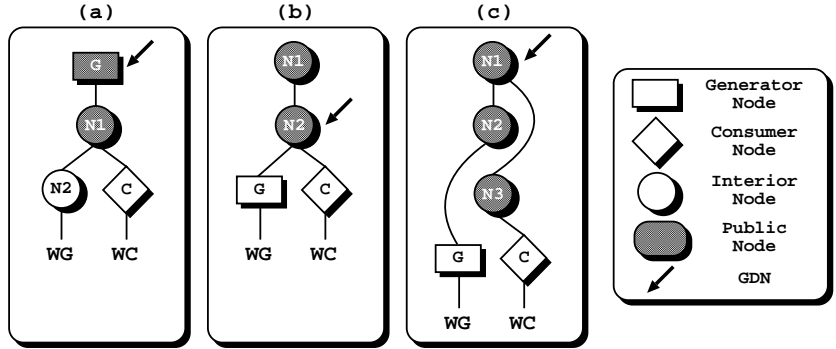}
}
\caption{Spotting the generator dependency node.}
\label{fig_public_generator_dependency}
\end{figure}

To clarify these new situations we introduce a new concept, the
\emph{Generator Dependency Node} (or \emph{GDN}). Its purpose is to
signal the nodes that are candidates to be leader nodes, therefore
representing a similar role as that of the generator nodes for
sequential tabling. A GDN is calculated whenever a new consumer node,
say ${\cal C}$, is created. We define the GDN ${\cal D}$ for a
consumer node ${\cal C}$ with generator $\cal G$ to be \emph{the
  youngest node on ${\cal C}$'s current branch that is an ancestor of
  ${\cal G}$}. Obviously, if ${\cal G}$ belongs to the current branch
of ${\cal C}$ then ${\cal G}$ must be the GDN. Thus GDN reduces to
leader node for sequential computations. On the other hand, if the
worker allocating ${\cal C}$ is not the one that allocated ${\cal G}$
then the youngest node ${\cal D}$ is a public node, but not
necessarily ${\cal G}$. Figure~\ref{fig_public_generator_dependency}
presents three different situations that better illustrate the GDN
concept. ${\cal W_G}$ is always the worker that allocated the
generator node ${\cal G}$, and ${\cal W_C}$ is the worker that is
allocating a consumer node ${\cal C}$.

In situation (a), the generator node ${\cal G}$ is on the branch of
the consumer node ${\cal C}$, and thus, ${\cal G}$ is the GDN. In
situation (b), nodes ${\cal N}_1$ and ${\cal N}_2$ are on the branch
of ${\cal C}$ and both contain a branch leading to the generator
${\cal G}$. As ${\cal N}_2$ is the youngest node of the two, it is the
GDN. Situation (c) differs from (b) in that the public nodes represent
more than one branch and, in this case, are interleaved in the
physical stack. In this situation, ${\cal N}_1$ is the unique node
that belongs to ${\cal C}$'s branch and that also contains ${\cal G}$
in a branch below. ${\cal N}_2$ contains ${\cal G}$ in a branch below,
but it is not on ${\cal C}$'s branch, while ${\cal N}_3$ is on ${\cal
  C}$'s branch, but it does not contain ${\cal G}$ in a branch below.
Therefore, ${\cal N}_1$ is the GDN. Notice that in both cases (b) and
(c) the GDN can be a generator, a consumer or an interior node.

The procedure that computes the leader node information when
allocating a new dependency frame now relies on the GDN
concept. Remember that it is through this information that a node can
determine whether it is a leader node. The main difference from the
sequential algorithm is that now we first hypothesize that the leader
node for the consumer node in hand is its GDN, and not its generator
node. Then, we check the consumer nodes younger than the newly found
GDN for an older dependency. Note that as soon as an older dependency
${\cal D}$ is found in a consumer node ${\cal C'}$, the remaining
consumer nodes, older than ${\cal C'}$ but younger than the GDN, do
not need to be checked. This is safe because the previous computation
of the leader node information for the consumer node ${\cal C'}$
already represents the oldest dependency that includes the remaining
consumer nodes. We next give an argument on the correctness of the
algorithm.

Consider a consumer node with GDN ${\cal G}$ and assume that its
leader node ${\cal D}$ is found in the dependency frame for consumer
node ${\cal C}$. Now hypothesize that there is a consumer node ${\cal
  N}$ younger than ${\cal G}$ with a reference ${\cal D'}$ older than
${\cal D}$. Therefore, when previously computing the leader node for
${\cal C}$ one of the following situations occurred: \textbf{(i)}
${\cal D}$ is the GDN for ${\cal C}$ or \textbf{(ii)} ${\cal D}$ was
found in a dependency frame for a consumer node ${\cal C'}$. Situation
\textbf{(i)} is not possible because ${\cal N}$ is younger than ${\cal
  D}$ and it holds a reference older than ${\cal D}$. Regarding
situation \textbf{(ii)}, ${\cal C'}$ is necessarily younger than
${\cal N}$ as otherwise the reference found for ${\cal C}$ had been
${\cal D'}$. By recursively applying the previous argument to the
computation of the leader node for ${\cal C'}$ we conclude that our
initial hypothesis cannot hold because the number of nodes between
${\cal C}$ and ${\cal N}$ is finite.

\begin{figure}[!ht]
\centerline{
\epsfxsize=9cm
\epsffile{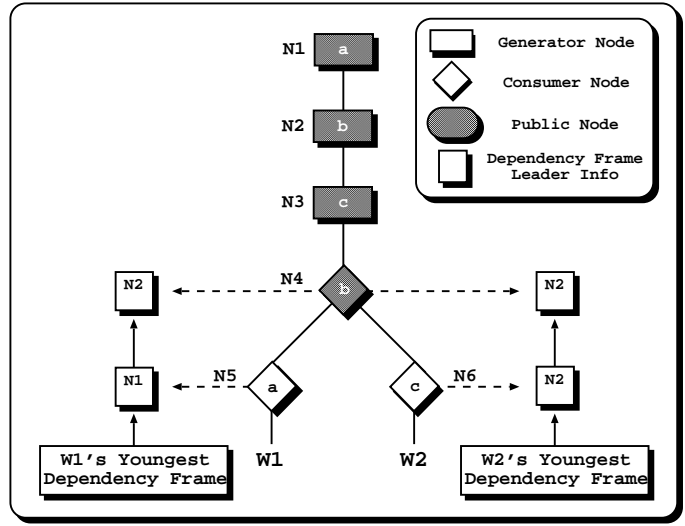}
}
\caption{Dependency frames in the parallel environment.}
\label{fig_dependency_frames}
\end{figure}

With this scheme, concurrency is not a problem. Each worker views its
own leader node independently from the execution being done by
others. A new consumer node is always a private node and a new
dependency frame is always the youngest dependency frame for a
worker. The leader information stored in a dependency frame denotes
the resulting leader node at the time the correspondent consumer node
was allocated. Thus, after computing such information it remains
unchanged. If when allocating a new consumer node the leader changes,
the new leader information is only stored in the dependency frame for
the new consumer, therefore not influencing others. Observe, for
example, the situation from Figure~\ref{fig_dependency_frames}. Two
workers, ${\cal W}_1$ and ${\cal W}_2$, exploiting different
alternatives from a common public node, ${\cal N}_4$, are allocating
new private consumer nodes. They compute the leader node information
for the new dependency frames without requiring any explicit
communication between both and without requiring any synchronization
if consulting the common dependency frame for node ${\cal N}_4$. The
resulting dependency chain for each worker is illustrated on each side
of the figure. Note that the dependency frame for consumer node ${\cal
N}_4$ is common to both workers. It is illustrated twice only for
simplicity.

Within this scenario, worker ${\cal W}_1$ will check for completion at
node ${\cal N}_1$, its current leader node, and worker ${\cal W}_2$
will check for completion at node ${\cal N}_2$. Obviously, ${\cal
  W}_2$ cannot perform completion when reaching ${\cal N}_2$. If
${\cal W}_1$ finds new answers for subgoal \texttt{c}, they should be
consumed in node ${\cal N}_6$. Moreover, as ${\cal W}_1$ has a
dependency for an older node, ${\cal N}_1$, the SCCs from both workers
should only be completed together at node ${\cal N}_1$. However,
${\cal W}_1$ can allocate another consumer node that changes its
current leader node. Therefore, ${\cal W}_2$ cannot know beforehand
the leader where both SCCs should be completed. Determining the leader
node where several dependent SCCs from different workers may be
completed together is the problem that we address next.

\subsection{SCC Suspension}
\label{section_scc}

Different paths may be followed when a worker ${\cal W}$ reaches a
leader node for a SCC ${\cal S}$. The simplest case is when the node
is private. In this case, we proceed as for sequential tabling.
Otherwise, the node is public, and other workers can still influence
${\cal S}$. For instance, these workers may find new answers for a
consumer node in ${\cal S}$, in which case the consumer must be
resumed to consume the new answers. Clearly, in such cases, ${\cal W}$
should not complete. On the other hand, ${\cal W}$ has tried all
available alternatives and would like to move anywhere in the tree,
say to node ${\cal N}$, to try other work. According to the copying
model we use for or-parallelism, we should backtrack to the youngest
node common to ${\cal N}$'s branch, that is, we should reset our
stacks to the values of the common node. According to the freezing
model that we use for tabling, we cannot recover the current consumers
because they are frozen. We thus have a contradiction.

Note that this is the only case where or-parallelism and tabling
conflict. One solution would be to disallow movement in this case.
Unfortunately, we would again severely restrict parallelism. As a
result, in order to allow ${\cal W}$ to continue execution it becomes
necessary to \emph{suspend the SCC} at hand. Suspending a SCC includes
saving the SCC's stacks to a proper space, leaving in the leader node
a reference to the suspended SCC. These suspended computations are
considered again when the remaining workers do completion.

In order to find out which suspended SCCs need to be resumed, each
worker maintains a list of nodes with suspended SCCs. The last worker
backtracking from a public node ${\cal N}$ checks if it holds
references to suspended SCCs. If so, then ${\cal N}$ is included in
the worker's list of nodes with suspended SCCs (the nodes are linked
in stack order). If the node already belongs to other worker's list,
it is not collected.

\begin{figure}[!ht]
\centerline{
\epsfxsize=12cm
\epsffile{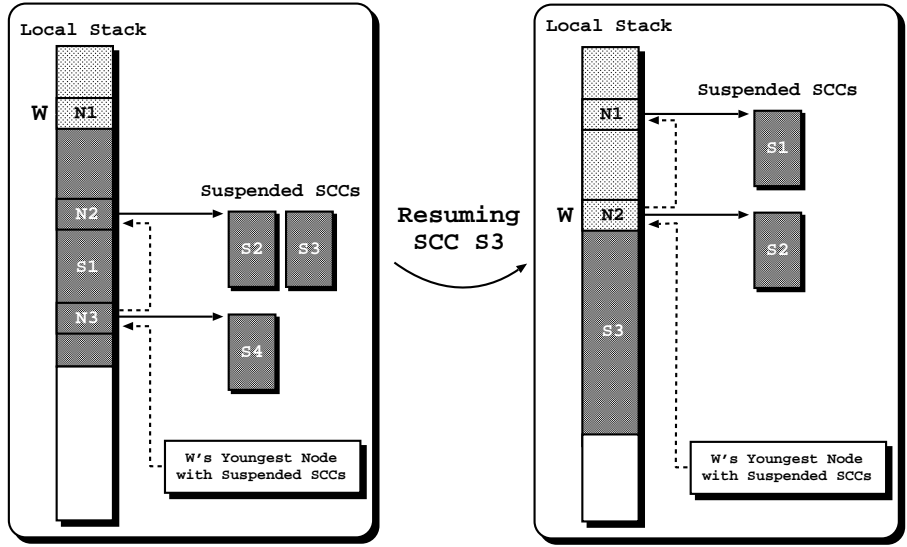}
}
\caption{Resuming a suspended SCC.}
\label{fig_resuming_scc}
\end{figure}

A suspended SCC should be resumed if it contains consumer nodes with
unconsumed answers. To resume a suspended SCC a worker needs to copy
the saved stacks to the correct position in its own stacks, and thus,
it has to suspend its current SCC first. Figure~\ref{fig_resuming_scc}
illustrates the management of suspended SCCs when searching for SCCs
to resume. It considers a worker ${\cal W}$, positioned in the leader
node ${\cal N}_1$ of its current SCC ${\cal S}_1$. ${\cal W}$ consults
its list of nodes with suspended SCCs, and starts checking the
suspended SCC ${\cal S}_4$ for unconsumed answers. Assuming that
${\cal S}_4$ does not contain unconsumed answers, the search continues
in the next node in the list. Here, suppose that SCC ${\cal S}_2$ does
not have consumer nodes with unconsumed answers, but SCC ${\cal S}_3$
does. The current SCC ${\cal S}_1$ is then suspended, and only then
${\cal S}_3$ resumed.

Notice that node ${\cal N}_3$ was removed from ${\cal W}$'s list of
suspended SCCs because ${\cal S}_3$ may not include ${\cal N}_3$ in
its stack segments. For simplicity and efficiency, instead of checking
${\cal S}_3$'s segments, we simply remove ${\cal N}_3$'s from ${\cal
  W}$'s list. Note that this is a safe decision as a SCC only depends
from branches below the leader node. Thus, if ${\cal S}_3$ does not
include ${\cal N}_3$ then no new answers can be found for ${\cal
  S}_4$'s consumer nodes. Otherwise, if this is not the case then
${\cal W}$ or other workers can eventually be scheduled to a node held
by ${\cal S}_4$ and find new answers for at least one of its consumer
nodes. In this case, when failing, these workers will necessarily
backtrack through ${\cal N}_3$, ${\cal S}_4$'s leader. Therefore, the
last worker backtracking from ${\cal N}_3$ will collect it for its own
list, which allows ${\cal S}_4$ to be later resumed when executing
completion in an older leader node.

\subsection{The Flow of Control}

\begin{figure}[!ht]
\centerline{
\epsfxsize=12cm
\epsffile{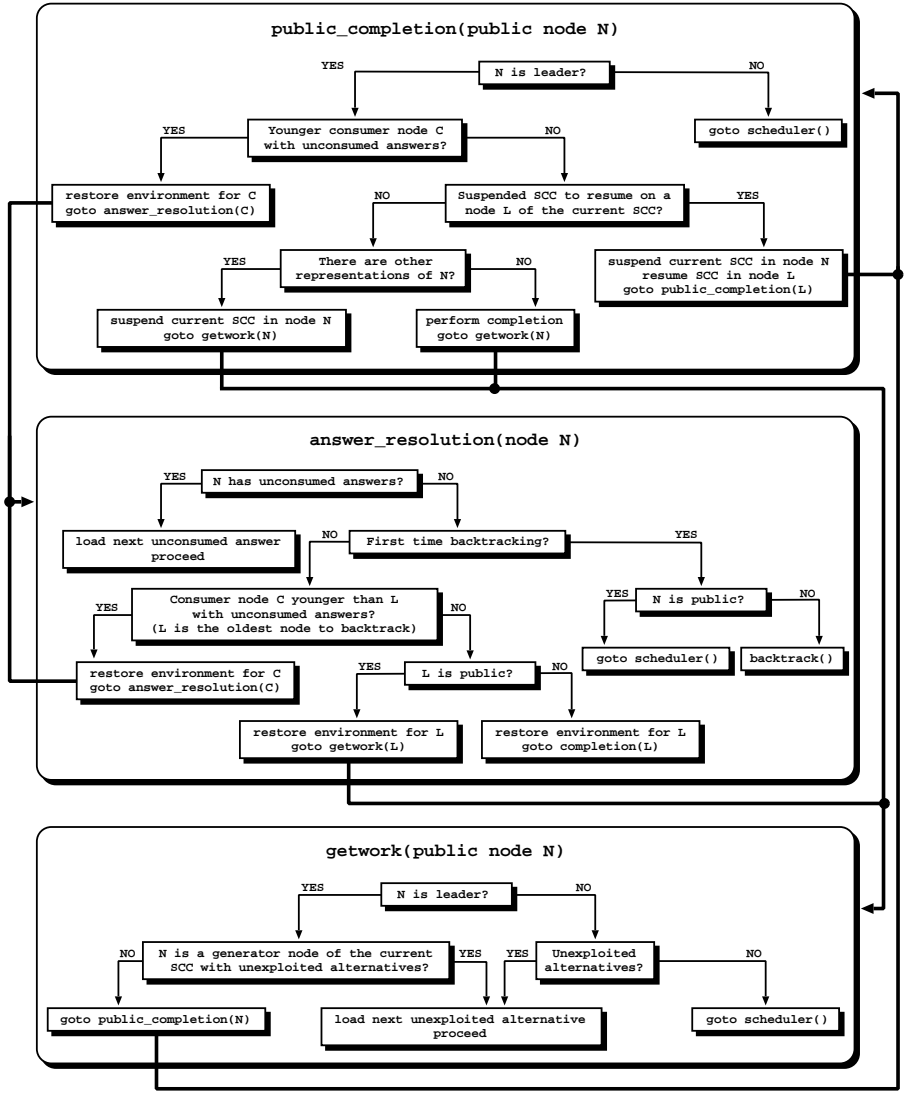}
}
\caption{The flow of control in a parallel tabled evaluation.}
\label{fig_control_flow}
\end{figure}

Actual execution control of a parallel tabled evaluation mainly flows
through four procedures. The process of completely evaluating SCCs is
accomplished by the \texttt{completion()} and
\texttt{answer\_resolution()} procedures, while parallel
synchronization is achieved by the \texttt{getwork()} and
\texttt{scheduler()} procedures. Here we focus on the execution in
engine mode, that is on the \texttt{completion()},
\texttt{answer\_resolution()} and \texttt{getwork()} procedures, and
leave scheduling for the following section.
Figure~\ref{fig_control_flow} presents a general overview of how
control flows between the three procedures and how it flows within
each procedure.

A novel completion procedure, \texttt{public\_completion()},
implements completion detection for public leader nodes. As for
private nodes, whenever a public node finds that it is a leader, it
starts to check for younger consumer nodes with unconsumed answers. If
there is such a node, we resume the computation to it. Otherwise, it
checks for suspended SCCs with unconsumed answers. Remember that to
resume a suspended SCC a worker needs to suspend its current SCC
first.

We thus adopted the strategy of resuming suspended SCCs \emph{only
  when the worker finds itself at a leader node}, since this is a
decision point where the worker either completes or suspends the
current SCC. Hence, if the worker resumes a suspended SCC it does not
introduce further dependencies. This is not the case if the worker
would resume a suspended SCC ${\cal R}$ as soon as it reached the node
where it had suspended. In that situation, the worker would have to
suspend its current SCC ${\cal S}$, and after resuming ${\cal R}$ it
would probably have to also resume ${\cal S}$ to continue its
execution. A first disadvantage is that the worker would have to make
more suspensions and resumptions. Moreover, if we resume earlier,
${\cal R}$ may include consumer nodes with unconsumed answers that are
common with ${\cal S}$. More importantly, suspending in non-leader
nodes leads to further complexity that can be very difficult to
manage.

A SCC ${\cal S}$ is completely evaluated when \textbf{(i)} there are
no unconsumed answers in any consumer node belonging to ${\cal S}$ or
in any consumer node within a SCC suspended in a node belonging to
${\cal S}$; and \textbf{(ii)} there are no other representations of
the leader node ${\cal N}$ in the computational environment, be ${\cal
N}$ represented in the execution stacks of a worker or be ${\cal N}$
in the suspended stack segments of a SCC. Completing a SCC includes
\textbf{(i)} marking all dependent subgoals as complete; \textbf{(ii)}
releasing the frames belonging to the complete branches, including the
branches in suspended SCCs; \textbf{(iii)} releasing the frozen stacks
and the memory space used to hold the stacks from suspended SCCs; and
\textbf{(iv)} readjusting the freeze registers and the whole set of
stack and frame pointers.

The answer resolution operation for the parallel environment
essentially uses the same algorithm as previously described for
private nodes (please refer to
section~\ref{section_completion_answer_resolution}). Initially, the
procedure checks for unconsumed answers to be loaded for execution. If
we have answers, execution will jump to them. Otherwise, we schedule
for a backtracking node. If this is not the first time that
backtracking from that consumer node takes place, we know that the
computation has been resumed from an older leader node ${\cal L}$
during an unsuccessful completion operation. ${\cal L}$ is thus the
oldest node to where we can backtrack. Backtracking must be done to
the next consumer node that has unconsumed answers and that is younger
than ${\cal L}$. Otherwise, if there are no such consumer nodes,
backtracking must be done to ${\cal L}$.

The \texttt{getwork()} procedure contributes to the progress of a
parallel tabled evaluation by moving to effective work. The usual way
to execute \texttt{getwork()} is through failure to the youngest
public node on the current branch. We can distinguish two main
procedures in \texttt{getwork()}. One detects completion points and
therefore makes the computation flow to the
\texttt{public\_completion()} procedure. The other corresponds to
or-parallel execution. It synchronizes to check for available
alternatives and executes the next one, if any. Otherwise, it invokes
the scheduler. A completion point is detected when ${\cal N}$ is the
leader node pointed by the youngest dependency frame. The exception is
if ${\cal N}$ is itself a generator node for a consumer node within
the current SCC and it contains unexploited alternatives. In such
cases, the current SCC is not fully exploited. Hence, we should
exploit first the available alternatives, and only then invoke
completion.

\subsection{Scheduling Work}

Scheduling work is the scheduler's task. It is about efficiently
distributing the available work for exploitation between the running
workers. In a parallel tabling environment we have the extra
constraint of keeping the correctness of sequential tabling semantics.
A worker enters in scheduling mode when it runs out of work and
returns to execution whenever a new piece of unexploited work is
assigned to it by the scheduler.

The scheduler for the OPTYap engine is mainly based on YapOr's
scheduler. All the scheduler strategies implemented for YapOr were
used in OPTYap. However, extensions were introduced in order to
preserve the correctness of tabling semantics. These extensions allow
support for leader nodes, frozen stack segments, and suspended
SCCs. The OPTYap model was designed to enclose the computation within
a SCC until the SCC was suspended or completely evaluated. Thus,
OPTYap introduces the constraint that the \emph{computation cannot
flow outside the current SCC, and workers cannot be scheduled to
execute at nodes older than their current leader node}. Therefore,
when scheduling for the nearest node with unexploited alternatives, if
it is found that the current leader node is younger than the potential
nearest node with unexploited alternatives, then the current leader
node is the node scheduled to proceed with the evaluation.

The next case is when the scheduling to determine the nearest node
with unexploited alternatives does not return any node to proceed
execution. The scheduler then starts searching for busy\footnote{A
worker is said to be busy when it is in engine mode exploiting
alternatives. A worker is said to be idle when it is in scheduling
mode searching for work.} workers that can be demanded for work. If
such a worker ${\cal B}$ is found, then the requesting worker moves up
to the youngest node that is common to ${\cal B}$, in order to become
partially consistent with part of ${\cal B}$. Otherwise, no busy
worker was found, and the scheduler moves the idle worker to a better
position in the search tree. Therefore, we can enumerate three
different situations for a worker to move up to a node ${\cal N}$:
\textbf{(i)} ${\cal N}$ is the nearest node with unexploited
alternatives; \textbf{(ii)} ${\cal N}$ is the youngest node common
with the busy worker we found; or \textbf{(iii)} ${\cal N}$
corresponds to a better position in the search tree.

The process of moving up in the search tree from a current node ${\cal
N}_0$ to a target node ${\cal N}_f$ is mainly implemented by the
\texttt{move\_up\_one\_node()} procedure. This procedure is invoked
for each node that has to be traversed until reaching ${\cal
N}_f$. The presence of frozen stack segments or the presence of
suspended SCCs in the nodes being traversed influences and can even
abort the usual moving up process.

Assume that the idle worker ${\cal W}$ is currently positioned at
${{\cal N}}_i$ and that it wants to move up one node. Initially, the
procedure checks for frozen nodes on the stack to infer whether ${\cal
W}$ is moving within a SCC. If so, ${\cal W}$ simply moves up. The
interesting case is when ${\cal W}$ is not within a SCC. If ${{\cal
N}}_i$ holds a suspended SCC, then ${\cal W}$ can safely resume it. If
resumption does not take place, the procedure proceeds to check
whether ${\cal W}$ holds the unique representation of ${\cal
N}_i$. This being the case, the suspended SCCs in ${\cal N}_i$ can be
completed. Completion can be safely performed over the suspended SCCs
in ${\cal N}_i$ not only because the SCCs are completely evaluated, as
none was previously resumed, but also because no more dependencies
exist, as there are no other branches below ${\cal N}_i$. Moreover, if
${\cal N}_i$ is a generator node then its correspondent subgoal can be
also marked as completed. Otherwise, ${\cal W}$ simply moves up.

The scheduler extensions described are mainly related with tabling
support. As the scheduling strategies inherited from the YapOr's
scheduler were designed for an or-parallel model, and not for an
or-parallel tabling model, further work is still needed to implement
and experiment with proper scheduling strategies that can take
advantage of the parallel tabling environment.

\subsection{Speculative Work}

In~\cite{Ciepielewski-91}, Ciepielewski defines speculative work as
\emph{work which would not be done in a system with one
processor}. The definition clearly shows that speculative work is an
implementation problem for parallelism and it must be addressed
carefully in order to reduce its impact. The presence of pruning
operators during or-parallel execution introduces the problem of
speculative work~\cite{Hausman-PhD,Ali-92a,Beaumont-93}. Prolog has an
explicit pruning operator, the \emph{cut} operator. When a computation
executes a cut operation, all branches to the right of the cut are
pruned. Computations that can potentially be pruned are thus
\emph{speculative}. Earlier execution of such computations may result
in wasted effort compared to sequential execution.

In parallel tabling, not only the answers found for the query goal may
not be valid, but also answers found for tabled predicates may be
invalidated. The problem here is even more serious because tabled
answers can be consumed elsewhere in the tree, which makes
impracticable any late attempt to prune computations resulting from
the consumption of invalid tabled answers. Indeed, consuming invalid
tabled answers may result in finding more invalid answers for the same
or other tabled predicates. Notice that finding and consuming answers
is the natural way to get a tabled computation going forward. Delaying
the consumption of answers may compromise such flow. Therefore, tabled
answers should be released as soon as it is found that they are safe
from being pruned. Whereas for all-solution queries the requirement is
that, at the end of the execution, we will have the set of valid
answers; in tabling the requirement is to have the set of valid tabled
answers released as soon as possible.

Currently, OPTYap implements an extension of the cut scheme proposed
by Ali and Karlsson~\cite{Ali-92a}, that prunes useless work as early
as possible, by optimizing the delivery of tabled answers as soon as
it is found that they are safe from being pruned~\cite{Rocha-PhD}. As
cut semantics for operations that prune tabled nodes is still an open
problem, OPTYap does not handle cut operations that prune tabled nodes
and for such cases execution is aborted.

\section{Related Work}

A first proposal on how to exploit implicit parallelism in tabling
systems was Freire's \emph{Table-parallelism}~\cite{Freire-95}. In
this model, each tabled subgoal is computed independently in a single
computational thread, a \emph{generator thread}. Each generator thread
is associated with a unique tabled subgoal and it is responsible for
fully exploiting its search tree in order to obtain the complete set
of answers. A generator thread dependent on other tabled subgoals will
asynchronously consume answers as the correspondent generator threads
will make them available. Within this model, parallelism results from
having several generator threads running concurrently. Parallelism
arising from non-tabled subgoals or from execution alternatives to
tabled subgoals is not exploited. Moreover, we expect that scheduling
and load balancing would be even harder than for traditional parallel
systems.

More recent work~\cite{Guo-01}, proposes a different approach to the
problem of exploiting implicit parallelism in tabled logic
programs. The approach is a consequence of a new sequential tabling
scheme based on \emph{dynamic reordering of alternatives with variant
calls}. This dynamic alternative reordering strategy not only tables
the answers to tabled subgoals, but also the alternatives leading to
variant calls, the \emph{looping alternatives}. Looping alternative
are reordered and placed at the end of the alternative list for the
call. After exploiting all matching clauses, the subgoal enters a
looping state, where the looping alternatives, if they exist, start
being tried repeatedly until a fixpoint is reached.  An important
characteristic of tabling is that it avoids recomputation of tabled
subgoals. An interesting point of the dynamic reordering strategy is
that it avoids recomputation through performing recomputation. The
process of retrying alternatives may cause redundant recomputations of
the non-tabled subgoals that appear in the body of a looping
alternative. It may also cause redundant consumption of answers if the
body of a looping alternative contains more than one variant subgoal
call. Within this model, parallelism arises if we schedule the
multiple looping alternatives to different workers. Therefore,
parallelism may not come so naturally as for SLD evaluations and
parallel execution may lead to doing more work.

There have been other proposals for concurrent tabling but in a
distributed memory context. Hu~\cite{Hu-PhD} was the first to
formulate a method for distributed tabled evaluation termed
\emph{Multi-Processor SLG (SLGMP)}. This method matches subgoals with
processors in a similar way to Freire's approach.  Each processor gets
a single subgoal and it is responsible for fully exploiting its search
tree and obtain the complete set of answers. One of the main
contributions of SLGMP is its controlled scheme of propagation of
subgoal dependencies in order to safely perform distributed
completion. An implementation prototype of SLGMP was developed, but as
far as we know no results have been reported.

A different approach for distributed tabling was proposed by
Damásio~\cite{Damasio-00}. The architecture for this proposal relies
on four types of components: a \emph{goal manager} that interfaces
with the outside world; a \emph{table manager} that selects the
clients for storing tables; \emph{table storage clients} that keep the
consumers and answers of tables; and \emph{prover clients} that
perform evaluation. An interesting aspect of this proposal is the
completion detection algorithm. It is based on a classical credit
recovery algorithm~\cite{Mattern-89} for distributed termination
detection. Dependencies among subgoals are not propagated and,
instead, a controller client, associated with each SCC, controls the
credits for its SCC and detects completion if the credits reach the
zero value. An implementation prototype has also been developed, but
further analysis is required.

Marques \emph{et al.}~\cite{Marques-00} have proposed an initial
design for an architecture for a multi-threaded tabling engine. Their
first aim is to implement an engine capable of processing multiple
query requests concurrently. The main idea behind this proposal seems
very interesting, however the work is still in an initial stage.

Other related mechanisms for sequential tabling have also been
proposed. Demoen and Sagonas proposed a copying approach to deal with
tabled evaluations and implemented two different models, the
CAT~\cite{Demoen-98} and the CHAT~\cite{Demoen-00}. The main idea of
the CAT implementation is that it replaces SLG-WAM's freezing of the
stacks by copying the state of suspended computations to a proper
separate stack area. The CHAT implementation improves the CAT design
by combining ideas from the SLG-WAM with those from the CAT. It avoids
copying all the execution stacks that represent the state of a
suspended computation by introducing a technique for freezing stacks
without using freeze registers.

Zhou \emph{et al.}~\cite{Zhou-00,Zhou-01a} developed a linear tabling
mechanism that works on a single SLD tree without requiring
suspensions/resumptions of computations. The main idea is to let
variant calls execute from the remaining clauses of the former first
call. It works as follows: when there are answers available in the
table, the call consumes the answers; otherwise, it uses the predicate
clauses to produce answers.  Meanwhile, if a call that is a variant of
some former call occurs, it takes the remaining clauses from the
former call and tries to produce new answers by using them. The
variant call is then repeatedly re-executed, until all the available
answers and clauses have been exhausted, that is, until a fixpoint is
reached.

\section{Performance Analysis}
\label{section_performance_analysis}

To assess the efficiency of our parallel tabling implementation and
address the question of whether parallel tabling is worthwhile, we
present next a detailed analysis of OPTYap's performance. We start by
presenting an overall view of the overheads of supporting the several Yap
extensions: YapOr, YapTab and OPTYap. Then, we compare YapOr's
parallel performance with that of OPTYap for a set of non-tabled
programs. Next, we use a set of tabled programs to measure the
sequential behavior of YapTab, OPTYap and XSB, and to assess OPTYap's
performance when running the tabled programs in parallel.

YapOr, YapTab and OPTYap are based on Yap's~4.2.1 engine\footnote{Note
that sequential execution would be somewhat better with more recent
Yap engines.}. We used the same compilation flags for Yap, YapOr,
YapTab and OPTYap. Regarding XSB Prolog, we used version~2.3 with the
default configuration and the default execution parameters. All
systems use batched scheduling for tabling.

The environment for our experiments was \emph{oscar}, a Silicon
Graphics Cray Origin2000 parallel computer from the Oxford
Supercomputing Centre. \emph{Oscar} consists of 96 MIPS 195 MHz R10000
processors each with 256 Mbytes of main memory (for a total shared
memory of 24 Gbytes) and running the IRIX~6.5.12 kernel. While
benchmarking, the jobs were submitted to an execution queue
responsible for scheduling the pending jobs through the available
processors in such a way that, when a job is scheduled for execution,
the processors attached to the job are fully available during the
period of time requested for execution. We have limited our
experiments to 32 processors because the machine was always with a
very high load and we were limited to a guest-account.

\subsection{Performance on Non-Tabled Programs}

Fundamental criteria to judge the success of an or-parallel, tabling,
or of a combined or-parallel tabling model includes measuring the
overhead introduced by the model when running programs that do not
take advantage of the particular extension. Ideally, a program should
not pay a penalty for mechanisms that it does not require.

To place our performance results in perspective we first evaluate how
the original Yap Prolog engine compares against the several Yap
extensions and against the most well-known tabling engine, XSB
Prolog. We use a set of standard non-tabled logic programming
benchmarks. All benchmarks find all the answers for the
problem. Multiple answers are computed through automatic failure after
a valid answer has been found. The set includes the following
benchmark programs:

\begin{description}
\item[cubes:] solves the N-cubes or instant insanity problem from
  Tick's book~\cite{Tick-91}. It consists of stacking 7 colored cubes
  in a column so that no color appears twice within any given side of
  the column.
  
\item[ham:] finds all hamiltonian cycles for a graph consisting of 26
  nodes with each node connected to other 3 nodes.
  
\item[map:] solves the problem of coloring a map of 10 countries with
  five colors such that no two adjacent countries have the same color.
  
\item[nsort:] naive sort algorithm. It sorts a list of 10 elements by
  brute force starting from the reverse order (and worst) case.
  
\item[puzzle:] places numbers 1 to 19 in an hexagon pattern such that
  the sums in all 15 diagonals add to the same value (also taken from
  Tick's book~\cite{Tick-91}).
  
\item[queens:] a non-naive algorithm to solve the problem of placing
  11 queens on a 11x11 chess board such that no two queens attack each
  other.
\end{description}

Table~\ref{non_tabled_sequential} shows the base execution time, in
seconds, for Yap, YapOr, YapTab, OPTYap and XSB for the set of
non-tabled benchmarks. In parentheses, it shows the overhead over the
Yap execution time. The timings reported for YapOr and OPTYap
correspond to the execution with a single worker. The results indicate
that YapOr, YapTab and OPTYap introduce, on average, an overhead of
about 10\%, 5\% and 17\% respectively over standard Yap. Regarding
XSB, the results show that, on average, XSB is 2.47 times slower than
Yap, a result mainly due to the faster Yap engine.

\begin{table}[!ht]
\caption{Yap, YapOr, YapTab, OPTYap and XSB execution time on non-tabled programs.}
\label{non_tabled_sequential}
\begin{tabular}{lrrrrr}
\hline\hline
    {\bf Bench}
    & \multicolumn{1}{c}{\bf Yap}
    & \multicolumn{1}{c}{\bf YapOr}
    & \multicolumn{1}{c}{\bf YapTab}
    & \multicolumn{1}{c}{\bf OPTYap}
    & \multicolumn{1}{c}{\bf XSB} \\
\hline
cubes       &  1.97 &  2.06 (1.05) &  2.05 (1.04) &  2.16 (1.10) &  4.81 (2.44) \\
ham         &  4.04 &  4.61 (1.14) &  4.28 (1.06) &  4.95 (1.23) & 10.36 (2.56) \\
map         &  9.01 & 10.25 (1.14) &  9.19 (1.02) & 11.08 (1.23) & 24.11 (2.68) \\
nsort       & 33.05 & 37.52 (1.14) & 35.85 (1.08) & 39.95 (1.21) & 83.72 (2.53) \\
puzzle      &  2.04 &  2.22 (1.09) &  2.19 (1.07) &  2.36 (1.16) &  4.97 (2.44) \\
queens      & 16.77 & 17.68 (1.05) & 17.58 (1.05) & 18.57 (1.11) & 36.40 (2.17) \\
\noalign{\vspace{.5cm}}
\multicolumn{2}{l}{\it Average}
                    &       (1.10) &       (1.05) &       (1.17) &       (2.47) \\
\hline\hline
\end{tabular}
\end{table}

YapOr overheads result from handling the work load register and from
testing operations that \textbf{(i)} verify whether a node is shared
or private, \textbf{(ii)} check for sharing requests, and
\textbf{(iii)} check for backtracking messages due to cut
operations. On the other hand, YapTab overheads are due to the
handling of the freeze registers and support of the forward
trail. OPTYap overheads inherits both sources of
overheads. Considering that Yap Prolog is one of the fastest Prolog
engines currently available, the low overheads achieved by YapOr,
YapTab and OPTYap are very good results.

Since OPTYap is based on the same environment model as the one used by
YapOr, we then compare OPTYap's performance with that of
YapOr. Table~\ref{non_tabled_parallel} shows the speedups relative to
the single worker case for YapOr and OPTYap with 4, 8, 16, 24 and 32
workers. Each speedup corresponds to the best execution time obtained
in a set of 3 runs. The results show that YapOr and OPTYap achieve
identical effective speedups in all benchmark programs. These results
allow us to conclude that OPTYap maintains YapOr's behavior in
exploiting or-parallelism in non-tabled programs, despite it including
all the machinery required to support tabled programs.

\begin{table}[!ht]
\caption{Speedups for YapOr and OPTYap on non-tabled programs.}
\label{non_tabled_parallel}
\begin{tabular}{lrrrrrrrrrrr}
\hline\hline
    & \multicolumn{5}{c}{\bf YapOr}
    &
    & \multicolumn{5}{c}{\bf OPTYap} \\ \noalign{\vspace{.2cm}}
    {\bf Bench}
    & \multicolumn{1}{c}{\bf 4}
    & \multicolumn{1}{c}{\bf 8}
    & \multicolumn{1}{c}{\bf 16}
    & \multicolumn{1}{c}{\bf 24}
    & \multicolumn{1}{c}{\bf 32}
    &
    & \multicolumn{1}{c}{\bf 4}
    & \multicolumn{1}{c}{\bf 8}
    & \multicolumn{1}{c}{\bf 16}
    & \multicolumn{1}{c}{\bf 24}
    & \multicolumn{1}{c}{\bf 32} \\
\hline
cubes   & 3.99 & 7.81 & 14.66 & 19.26 & 20.55 & & 3.98 & 7.74 & 14.29 & 18.67 & 20.97 \\
ham     & 3.93 & 7.61 & 13.71 & 15.62 & 15.75 & & 3.92 & 7.64 & 13.54 & 16.25 & 17.51 \\
map     & 3.98 & 7.73 & 14.03 & 17.11 & 18.28 & & 3.98 & 7.88 & 13.74 & 18.36 & 16.68 \\
nsort   & 3.98 & 7.92 & 15.62 & 22.90 & 29.73 & & 3.96 & 7.84 & 15.50 & 22.75 & 29.47 \\
puzzle  & 3.93 & 7.56 & 13.71 & 18.18 & 16.53 & & 3.93 & 7.51 & 13.53 & 16.57 & 16.73 \\
queens  & 4.00 & 7.95 & 15.39 & 21.69 & 25.69 & & 3.99 & 7.93 & 15.41 & 20.90 & 25.23 \\
\noalign{\vspace{.5cm}}
{\it Average}
        & 3.97 & 7.76 & 14.52 & 19.13 & 21.09 & & 3.96 & 7.76 & 14.34 & 18.92 & 21.10 \\
\hline\hline
\end{tabular}
\end{table}

\subsection{Performance on Tabled Programs}

In order to place OPTYap's results in perspective we start by
analyzing the overheads introduced to extend YapTab to parallel
execution and by measuring YapTab and OPTYap behavior when compared
with XSB. We use a set of tabled benchmark programs from the
XMC\footnote{The XMC system~\cite{Ramakrishnan-00} is a model checker
implemented atop the XSB system which verifies properties written in
the alternation-free fragment of the modal
$\mu$-calculus~\cite{Kozen-83} for systems specified in XL, an
extension of value-passing CCS~\cite{Milner-89}.}~\cite{xmc} and
XSB~\cite{xsb} \emph{world wide web} sites that are frequently used in
the literature to evaluate such systems. The benchmark programs are:

\begin{description}
\item[sieve:] the transition relation graph for the \emph{sieve}
  specification\footnote{We are thankful to C. R. Ramakrishnan for
    helping us in dumping the transition relation graph of the
    automatons corresponding to each given XL specification, and in
    building runnable versions out of the XMC environment.} defined
  for 5 processes and 4 overflow prime numbers.

\item[leader:] the transition relation graph for the \emph{leader
    election} specification defined for 5 processes.
  
\item[iproto:] the transition relation graph for the \emph{i-protocol}
  specification defined for a correct version (fix) with a huge window
  size (w = 2).
  
\item[samegen:] solves the same generation problem for a randomly
  generated 24x24x2 cylinder. This benchmark is very interesting
  because for sequential execution it does not allocate any consumer
  node. Variant calls to tabled subgoals only occur when the subgoals
  are already completed.
  
\item[lgrid:] computes the transitive closure of a 25x25 grid using a
  left recursion algorithm. A link between two nodes, $n$ and $m$, is
  defined by two different relations; one indicates that we can reach
  $m$ from $n$ and the other indicates that we can reach $n$ from $m$.
  
\item[lgrid/2:] the same as \textbf{lgrid} but it only requires half
  the relations to indicate that two nodes are connected. It defines
  links between two nodes by a single relation, and it uses a
  predicate to achieve symmetric reachability. This modification
  alters the order by which answers are found. Moreover, as indexing
  in the first argument is not possible for some calls, the execution
  time increases significantly. For this reason, we only use here a
  20x20 grid.

\item[rgrid/2:] the same as \textbf{lgrid/2} but it computes the
  transitive closure of a 25x25 grid and it uses a right recursion
  algorithm.
\end{description}

Table~\ref{tabled_sequential} shows the execution time, in seconds,
for YapTab, OPTYap and XSB for the set of tabled benchmarks. In
parentheses, it shows the overhead over the YapTab execution time. The
execution time reported for OPTYap correspond to the execution with a
single worker.

\begin{table}[!ht]
\caption{YapTab, OPTYap and XSB execution time on tabled programs.}
\label{tabled_sequential}
\begin{tabular}{lrrr}
\hline\hline
    {\bf Bench}
    & \multicolumn{1}{c}{\bf YapTab}
    & \multicolumn{1}{c}{\bf OPTYap}
    & \multicolumn{1}{c}{\bf XSB} \\
\hline
sieve   & 235.31 & 268.13 (1.14) & 433.53 (1.84) \\
leader  &  76.60 &  85.56 (1.12) & 158.23 (2.07) \\
iproto  &  20.73 &  23.68 (1.14) &  53.04 (2.56) \\
samegen &  23.36 &  26.00 (1.11) &  37.91 (1.62) \\
lgrid   &   3.55 &   4.28 (1.21) &   7.41 (2.09) \\
lgrid/2 &  59.53 &  69.02 (1.16) &  98.22 (1.65) \\
rgrid/2 &   6.24 &   7.51 (1.20) &  15.40 (2.47) \\
\noalign{\vspace{.5cm}}
{\it Average} &  &        (1.15) &        (2.04) \\
\hline\hline
\end{tabular}
\end{table}

The results indicate that, for these set of tabled benchmark programs,
OPTYap introduces, on average, an overhead of about 15\% over
YapTab. This overhead is very close to that observed for non-tabled
programs (11\%). The small difference results from locking requests to
handle the data structures introduced by tabling.  Locks are require
to insert new trie nodes into the table space, and to update subgoal
and dependency frame pointers to tabled answers. These locking
operations are all related with the management of tabled
answers. Therefore, the benchmarks that deal with more tabled answers
are the ones that potentially can perform more locking
operations. This causal relation seems to be reflected in the
execution times showed in Table~\ref{tabled_sequential}, because the
benchmarks that show higher overheads are also the ones that find more
answers. The answers found by each benchmark are presented next in
Table~\ref{tabled_stats}.

Table~\ref{tabled_sequential} also shows that YapTab is on average
about twice as fast as XSB for these set of benchmarks. This may be
partly due to the faster Yap engine, as seen in
Table~\ref{non_tabled_sequential}, and also to the fact that XSB
implements functionalities that are still lacking in YapTab and that
XSB may incur overheads in supporting those functionalities. These
results show that we have accomplished our initial aim of implementing
an or-parallel tabling system that compares favorably with current
\emph{state of the art} technology.  Hence, we believe the following
evaluation of the parallel engine is significant and fair.

In order to achieve a deeper insight on the behavior of each
benchmark, and therefore clarify some of the results presented next,
we first present in Table~\ref{tabled_stats} data on the benchmark
programs. The columns in Table~\ref{tabled_stats} have the following
meaning:

\begin{description} 
\item[first:] is the number of first calls to subgoals corresponding
  to tabled predicates. It corresponds to the number of generator
  choice points allocated.
  
\item[nodes:] is the number of subgoal/answer trie nodes used to
  represent the complete subgoal/answer trie structures of the tabled
  predicates in the given benchmark. For the answer tries, in
  parentheses, it shows the percentage of saving that the trie's
  design achieves on these data structures. Given the $total$ number
  of nodes required to represent individually each answer and the
  number of nodes $used$ by the trie structure, the $saving$ can be
  obtained by the following expression:
\[saving~=~\frac{total~-~used}{total}\]
As an example, consider two answers whose single representation
requires respectively 12 and 8 answer trie nodes for each. Assuming
that the answer trie representation of both answers only requires 15
answer trie nodes, thus 5 of those being common to both paths, it
achieves a saving of 25\%. Higher percentages of saving reflect higher
probabilities of lock contention when concurrently accessing the table
space.

\item[depth:] is the average depth of the whole set of paths in the
  corresponding answer trie structure. In other words, it is the
  average number of answer trie nodes required to represent an answer.
  Trie structures with smaller average depth values are more amenable
  to higher lock contention.
  
\item[unique:] is the number of non-redundant answers found for tabled
  subgoals. It corresponds to the number of answers stored in the
  table space.
  
\item[repeated:] is the number of redundant answers found for tabled
  subgoals. A high number of redundant answers can degrade the
  performance of the parallel system when using table locking schemes
  that lock the table space without taking into account whether
  writing to the table is, or is not, likely.
\end{description}

\begin{table}[!ht]
\caption{Characteristics of the tabled programs.}
\label{tabled_stats}
\begin{tabular}{lrrrrrrrr}
\hline\hline
    & \multicolumn{2}{c}{\bf Subgoal Tries}
    &
    & \multicolumn{2}{c}{\bf Answer Tries}
    &
    & \multicolumn{2}{c}{\bf New Answers} \\ \noalign{\vspace{.2cm}}
      {\bf Bench}
    & \multicolumn{1}{c}{\bf first}
    & \multicolumn{1}{c}{\bf nodes}
    &
    & \multicolumn{1}{c}{\bf nodes}
    & \multicolumn{1}{c}{\bf depth}
    &
    & \multicolumn{1}{c}{\bf unique}
    & \multicolumn{1}{c}{\bf repeated} \\
\hline
sieve   &   1 &    7 & &    8624(57\%) & 53   & &    380 & 1386181 \\
leader  &   1 &    5 & &   41793(70\%) & 81   & &   1728 &  574786 \\
iproto  &   1 &    6 & & 1554896(77\%) & 51   & & 134361 &  385423 \\
samegen & 485 &  971 & &   24190(33\%) &  1.5 & &  23152 &   65597 \\
lgrid   &   1 &    3 & &  391251(49\%) &  2   & & 390625 & 1111775 \\
lgrid/2 &   1 &    3 & &  160401(49\%) &  2   & & 160000 &  449520 \\
rgrid/2 & 626 & 1253 & &  782501(33\%) &  1.5 & & 781250 & 2223550 \\
\hline\hline
\end{tabular}
\end{table}

By observing Table~\ref{tabled_stats} it seems that \emph{sieve} and
\emph{leader} are the benchmarks least amenable to table lock
contention because they are the ones that find the least number of
answers and also the ones that have the deepest trie structures. In
this regard, \emph{lgrid}, \emph{lgrid/2} and \emph{rgrid/2}
correspond to the opposite case. They find the largest number of
answers and they have very shallow trie structures. However,
\emph{rgrid/2} is a benchmark with a large number of first subgoals
calls which can reduce the probability of lock contention because
answers can be found for different subgoal calls and therefore be
inserted with minimum overlap.  Likewise, \emph{samegen} is a
benchmark that can also benefit from its large number of first subgoal
calls, despite also presenting a very shallow trie structure.
Finally, \emph{iproto} is a benchmark that can also lead to higher
ratios of lock contention. It presents a deep trie structure, but it
inserts a huge number of trie nodes in the table space. Moreover, it
is the benchmark showing the highest percentage of saving.

To assess OPTYap's performance when running tabled programs in
parallel, we ran OPTYap with varying number of workers for the set of
tabled benchmark programs. Table~\ref{tabled_parallel_batched}
presents the speedups for OPTYap with 4, 8, 16, 24 and 32 workers. The
speedups are relative to the single worker case of
Table~\ref{tabled_sequential}. They correspond to the best speedup
obtained in a set of 3 runs. The table is divided in two main blocks:
the upper block groups the benchmarks that showed potential for
parallel execution, whilst the bottom block groups the benchmarks that
do not show any gains when run in parallel.

\begin{table}[!ht]
\caption{Speedups for OPTYap on tabled programs.}
\label{tabled_parallel_batched}
\begin{tabular}{lrrrrr}
\hline\hline
    & \multicolumn{5}{c}{\bf Number of Workers} \\ \noalign{\vspace{.2cm}}
    {\bf Bench}
    & \multicolumn{1}{c}{\bf  4}
    & \multicolumn{1}{c}{\bf  8}
    & \multicolumn{1}{c}{\bf 16}
    & \multicolumn{1}{c}{\bf 24}
    & \multicolumn{1}{c}{\bf 32} \\
\hline
sieve   & 3.99 & 7.97 & 15.87 & 23.78 & 31.50 \\
leader  & 3.98 & 7.92 & 15.78 & 23.57 & 31.18 \\
iproto  & 3.05 & 5.08 &  9.01 &  8.81 &  7.21 \\
samegen & 3.72 & 7.27 & 13.91 & 19.77 & 24.17 \\
lgrid/2 & 3.63 & 7.19 & 13.53 & 19.93 & 24.35 \\
\noalign{\vspace{.5cm}}
{\it Average}
        & 3.67 & 7.09 & 13.62 & 19.17 & 23.68 \\
\hline
lgrid   & 0.65 & 0.68 &  0.55 &  0.46 &  0.39 \\
rgrid/2 & 0.94 & 1.15 &  0.72 &  0.77 &  0.65 \\
\noalign{\vspace{.5cm}}
{\it Average}
        & 0.80 & 0.92 &  0.64 &  0.62 &  0.52 \\
\hline\hline
\end{tabular}
\end{table}

The results show superb speedups for the XMC \emph{sieve} and the
\emph{leader} benchmarks up to 32 workers. These benchmarks reach
speedups of 31.5 and 31.18 with 32 workers! Two other benchmarks in
the upper block, \emph{samegen} and \emph{lgrid/2}, also show
excellent speedups up to 32 workers. Both reach a speedup of 24 with
32 workers. The remaining benchmark, \emph{iproto}, shows a good
result up to 16 workers and then it slows down with 24 and 32 workers.
Globally, the results for the upper block are quite good, especially
considering that they include the three XMC benchmarks that are more
representative of real-world applications.

On the other hand, the bottom block shows almost no speedups at all.
Only for \emph{rgrid/2} with 8 workers we obtain a slight positive
speedup of 1.15. The worst case is for \emph{lgrid} with 32 workers,
where we are about 2.5 times slower than execution with a single
worker. In this case, surprisingly, we observed that for the whole set
of benchmarks the workers are busy for more than 95\% of the execution
time, even for 32 workers. The actual slowdown is therefore not caused
because workers became idle and start searching for work, as usually
happens with parallel execution of non-tabled programs. Here the
problem seems more complex: workers do have available work, but there
is a lot of contention to access that work.

The parallel execution behavior of each benchmark program can be
better understood through the statistics described in the tables that
follows. The columns in these tables have the following meaning:

\begin{description}
\item[variant:] is the number of variant calls to subgoals
  corresponding to tabled predicates. It matches the number of
  consumer choice points allocated.
  
\item[complete:] is the number of variant calls to completed tabled
  subgoals. It is when the \emph{completed table optimization} takes
  places, that is, when the set of found answers is consumed by
  executing compiled code directly from the trie structure associated
  with the completed subgoal.
  
\item[SCC suspend:] is the number of SCCs suspended.
  
\item[SCC resume:] is the number of suspended SCCs that were resumed.
  
\item[contention points:] is the total number of unsuccessful first
  attempts to lock data structures of all types. Note that when a
  first attempt fails, the requesting worker performs arbitrarily
  locking requests until it succeeds. Here, we only consider the first
  attempts.

  \begin{description}
  \item[subgoal frame:] is the number of unsuccessful first attempts
    to lock subgoal frames. A subgoal frame is locked in three main
    different situations: \textbf{(i)} when a new answer is found
    which requires updating the subgoal frame pointer to the last
    found answer; \textbf{(ii)} when marking a subgoal as completed;
    \textbf{(iii)} when traversing the whole answer trie structure to
    remove pruned answers and compute the code for direct compiled
    code execution.
    
  \item[dependency frame:] is the number of unsuccessful first
    attempts to lock dependency frames. A dependency frame has to be
    locked when it is checked for unconsumed answers.
    
  \item[trie node:] is the number of unsuccessful first attempts to
    lock trie nodes. Trie nodes must be locked when a worker has to
    traverse the subgoal trie structure during a tabled subgoal call
    operation or the answer trie structure during a new answer
    operation.
  \end{description}
\end{description}

To accomplish these statistics it was necessary to introduce in the
system a set of counters to measure the several parameters. Although,
the counting mechanism introduces an additional overhead in the
execution time, we assume that it does not significantly influence the
parallel execution pattern of each benchmark program.

Tables~\ref{stats_batched_upper} and~\ref{stats_batched_below} show
respectively the statistics gathered for the group of programs with
and without parallelism. We do not include the statistics for the
\emph{leader} benchmark because its execution behavior showed to be
identical to the observed for the \emph{sieve} benchmark.

\begin{table}[!ht]
\caption{Statistics of OPTYap using batched scheduling for the group
         of programs with parallelism.}
\label{stats_batched_upper}
\begin{tabular}{lrrrrr}
\hline\hline
    & \multicolumn{5}{c}{\bf Number of Workers} \\ \noalign{\vspace{.2cm}}
    {\bf Parameter}
    & \multicolumn{1}{c}{\bf  4}
    & \multicolumn{1}{c}{\bf  8}
    & \multicolumn{1}{c}{\bf 16}
    & \multicolumn{1}{c}{\bf 24}
    & \multicolumn{1}{c}{\bf 32} \\
\hline
{\bf sieve}         &        &        &        &        &        \\
variant/complete    &    1/0 &    1/0 &    1/0 &    1/0 &    1/0 \\
SCC suspend/resume  &   20/0 &   70/0 &  136/0 &  214/0 &  261/0 \\
contention points   &    108 &    329 &    852 &   1616 &   3040 \\
~~~subgoal frame    &      0 &      0 &      0 &      0 &      2 \\
~~~dependency frame &      0 &      0 &      1 &      0 &      4 \\
~~~trie node        &     96 &    188 &    415 &    677 &   1979 \\
\noalign{\vspace{.5cm}}
{\bf iproto}        &        &        &        &        &        \\
variant/complete    &    1/0 &    1/0 &    1/0 &    1/0 &    1/0 \\
SCC suspend/resume  &    5/0 &    9/0 &   17/0 &   26/0 &   32/0 \\
contention points   &   7712 &  22473 &  60703 & 120162 & 136734 \\
~~~subgoal frame    &   3832 &   9894 &  21271 &  33162 &  33307 \\
~~~dependency frame &    678 &   4685 &  25006 &  66334 &  81515 \\
~~~trie node        &   3045 &   6579 &  10537 &  11816 &  11736 \\
\noalign{\vspace{.5cm}}
{\bf samegen}       &          &          &          &          &          \\
variant/complete    & 485/1067 & 1359/193 & 1355/197 & 1384/168 & 1363/189 \\
SCC suspend/resume  &    187/2 &   991/11 &  1002/20 &  1024/25 &  1020/34 \\
contention points   &      255 &      314 &      743 &     1160 &     1607 \\
~~~subgoal frame    &        8 &       52 &      112 &      283 &      493 \\
~~~dependency frame &        0 &        0 &        1 &        0 &        0 \\
~~~trie node        &      154 &      119 &      201 &      364 &      417 \\
\noalign{\vspace{.5cm}}
{\bf lgrid/2}       &        &        &        &        &        \\
variant/complete    &    1/0 &    1/0 &    1/0 &    1/0 &    1/0 \\
SCC suspend/resume  &    4/0 &    8/0 &   16/0 &   24/0 &   32/0 \\
contention points   &   4004 &  10072 &  28669 &  59283 &  88541 \\
~~~subgoal frame    &    167 &   1124 &   7319 &  17440 &  27834 \\
~~~dependency frame &     98 &   1209 &   5987 &  23357 &  35991 \\
~~~trie node        &   2958 &   5292 &  10341 &  12870 &  12925 \\
\hline\hline
\end{tabular}
\end{table}

The statistics obtained for the \emph{sieve} benchmark support the
excellent performance speedups showed for parallel execution. It shows
insignificant number of contention points, it only calls a variant
subgoal, and despite the fact that it suspends some SCCs it
successfully avoids resuming them. In this regard, the \emph{samegen}
benchmark also shows insignificant number of contention points.
However the number of variant subgoals calls and the number of
suspended/resumed SCCs indicate that it introduces more dependencies
between workers. Curiously, for more than 4 workers, the number of
variant calls and the number of suspended SCCs seems to be stable. The
only parameter that slightly increases is the number of resumed SCCs.
Regarding \emph{iproto} and \emph{lgrid/2}, lock contention seems to
be the major problem. Trie nodes show identical lock contention,
however \emph{iproto} inserts about 10 times more answer trie nodes
than \emph{lgrid/2}. Subgoal and dependency frames show an identical
pattern of contention, but \emph{iproto} presents higher contention
ratios. Moreover, if we remember from Table~\ref{tabled_sequential}
that \emph{iproto} is about 3 times faster than \emph{lgrid/2} to
execute, we can conclude that the contention ratio for \emph{iproto}
is obviously much higher per time unit, which justifies its worst
behavior.

\begin{table}[!ht]
\caption{Statistics of OPTYap using batched scheduling for the group
         of programs without parallelism.}
\label{stats_batched_below}
\begin{tabular}{lrrrrr}
\hline\hline
    & \multicolumn{5}{c}{\bf Number of Workers} \\ \noalign{\vspace{.2cm}}
    {\bf Parameter}
    & \multicolumn{1}{c}{\bf  4}
    & \multicolumn{1}{c}{\bf  8}
    & \multicolumn{1}{c}{\bf 16}
    & \multicolumn{1}{c}{\bf 24}
    & \multicolumn{1}{c}{\bf 32} \\
\hline
{\bf lgrid}         &        &        &        &        &        \\
variant/complete    &    1/0 &    1/0 &    1/0 &    1/0 &    1/0 \\
SCC suspend/resume  &    4/0 &    8/0 &   16/0 &   24/0 &   32/0 \\
contention points   & 112740 & 293328 & 370540 & 373910 & 452712 \\
~~~subgoal frame    &  18502 &  73966 &  77930 &  68313 & 115862 \\
~~~dependency frame &  17687 & 113594 & 215429 & 223792 & 248603 \\
~~~trie node        &  72751 &  91909 &  61857 &  62629 &  64029 \\
\noalign{\vspace{.5cm}}
{\bf rgrid/2}       &           &           &           &          &           \\
variant/complete    & 3051/1124 & 3072/1103 & 3168/1007 & 3226/949 &  3234/941 \\
SCC suspend/resume  &  1668/465 &  1978/766 & 2326/1107 & 2121/882 & 2340/1078 \\
contention points   &     58761 &    110984 &    133058 &   170653 &    173773 \\
~~~subgoal frame    &     55415 &    103104 &    122938 &   159709 &    160771 \\
~~~dependency frame &         0 &         8 &         5 &      259 &       268 \\
~~~trie node        &      1519 &      3595 &      5016 &     4780 &      4737 \\
\hline\hline
\end{tabular}
\end{table}

The statistics gathered for the second group of programs present very
interesting results. Remember that \emph{lgrid} and \emph{rgrid/2} are
the benchmarks that find the largest number of answers per time unit
(please refer to Tables~\ref{tabled_sequential}
and~\ref{tabled_stats}). Regarding \emph{lgrid}'s statistics it shows
high contention ratios in all parameters considered. Closer analysis
of its statistics allows us to observe that it shows an identical
pattern when compared with \emph{lgrid/2}. The problem is that the
ratio per time unit is significantly worst for \emph{lgrid}. This
reflects the fact that most of \emph{lgrid}'s execution time is spent
in \emph{massively} accessing the table space to insert new answers
and to consume found answers.

The sequential order by which answers are accessed in the trie
structure is the key issue that reflects the high number of contention
points in subgoal and dependency frames. When inserting a new answer
we need to update the subgoal frame pointer to point at the last found
answer. When consuming a new answer we need to update the dependency
frame pointer to point at the last consumed answer. For programs that
find a large number of answers per time unit, this obviously increases
contention when accessing such pointers. Regarding trie nodes, the
small depth of \emph{lgrid}'s answer trie structure (2 trie nodes) is
one of the main factors that contributes to the high number of
contention points when massively inserting trie nodes. Trie structures
are a compact data structure. Therefore, obtaining good parallel
performance in the presence of massive table access will always be a
difficult task.

Analyzing the statistics for \emph{rgrid/2}, the number of variant
subgoals calls and the number of suspended/resumed SCCs suggest that
this benchmark leads to complex dependencies between workers.
Curiously, despite the large number of consumer nodes that the
benchmark allocates, contention in dependency frames is not a problem.
On the other hand, contention for subgoal frames seems to be a major
problem. The statistics suggest that the large number of SCC resume
operations and the large number of answers that the benchmark finds
are the key aspects that constrain parallel performance. A closer
analysis shows that the number of resumed SCCs is approximately
constant with the increase in the number of workers. This may suggest
that there are answers that can only be found when other answers are
also found, and that the process of finding such answers cannot be
anticipated. In consequence, suspended SCCs have always to be resumed
to consume the answers that cannot be found sooner. We believe that
the sequencing in the order that answers are found is the other major
problem that restrict parallelism in tabled programs.

Another aspect that can negatively influence this benchmark is the
number of completed calls. Before executing the first call to a
completed subgoal we need to traverse the trie structure of the
completed subgoal. When traversing the trie structure the
correspondent subgoal frame is locked. As \emph{rgrid/2} stores a huge
number of answer trie nodes in the table (please refer to
Table~\ref{tabled_stats}) this can lead to longer periods of lock
contention.

\section{Concluding Remarks}

We have presented the design, implementation and evaluation of
OPTYap. OPTYap is the first available system that exploits
or-parallelism and tabling from logic programs. A major guideline for
OPTYap was concerned with making best use of the excellent technology
already developed for previous systems. In this regard, OPTYap uses
Yap's efficient sequential Prolog engine as its starting framework,
and the SLG-WAM and environment copying approaches, respectively, as
the basis for its tabling and or-parallel components.

Through this research we aimed at showing that the models developed to
exploit implicit or-parallelism in standard logic programming systems
can also be used to successfully exploit implicit or-parallelism in
tabled logic programming systems. First results reinforced our belief
that tabling and parallelism are a very good match that can contribute
to expand the range of applications for Logic Programming.

OPTYap introduces low overheads for sequential execution and compares
favorably with current versions of XSB. Moreover, it maintains YapOr's
effective speedups in exploiting or-parallelism in non-tabled
programs.  Our best results for parallel execution of tabled programs
were obtained on applications that have a limited number of tabled
nodes, but high or-parallelism. However, we have also obtained good
speedups on applications with a large number of tabled nodes.

On the other hand, there are tabled programs where OPTYap may not
speed up execution. Table access has been the main factor limiting
parallel speedups so far. OPTYap implements tables as tries, thus
obtaining good indexing and compression. On the other hand, tries are
designed to avoid redundancy. To do so, they restrict concurrency,
especially when updating. We plan to study whether alternative designs
for the table data structure can obtain scalable speedups even when
frequently updating tables.

Our applications do not show the completion algorithm to be a major
factor in performance so far. In the future, we plan to study OPTYap
over a large range of applications, namely, natural language, database
processing, and non-monotonic reasoning. We expect that non-monotonic
reasoning applications, for instance, will raise more complex
dependencies and further stress the completion algorithm. We are also
interested in the implementation of pruning in the parallel
environment.

\section*{Acknowledgments}

The authors are thankful to the anonymous reviewers for their valuable
comments. This work has been partially supported by $CLoP^n$ (CNPq),
PLAG (FAPERJ), APRIL (POSI/SRI/40749/2001), and by funds granted to
LIACC through the Programa de Financiamento Plurianual, Funda\c{c}\~ao
para a Ci\^encia e Tecnologia and Programa POSI.

\bibliographystyle{plain}
\bibliography{references}

\begin{thebibliography}{10}

\bibitem{Ali-90b}
K.~Ali and R.~Karlsson.
\newblock {Full Prolog and Scheduling OR-Parallelism in Muse}.
\newblock {\em International Journal of Parallel Programming}, 19(6):445--475,
  1990.

\bibitem{Ali-90a}
K.~Ali and R.~Karlsson.
\newblock {The Muse Approach to OR-Parallel Prolog}.
\newblock {\em International Journal of Parallel Programming}, 19(2):129--162,
  1990.

\bibitem{Ali-92a}
K.~Ali and R.~Karlsson.
\newblock {Scheduling Speculative Work in MUSE and Performance Results}.
\newblock {\em International Journal of Parallel Programming}, 21(6):449--476,
  1992.

\bibitem{Apt-94}
K.~Apt and R.~Bol.
\newblock {Logic Programming and Negation: A Survey}.
\newblock {\em Journal of Logic Programming}, 19 \& 20:9--72, 1994.

\bibitem{Beaumont-93}
A.~Beaumont and D.~H.~D. Warren.
\newblock {Scheduling Speculative Work in Or-Parallel Prolog Systems}.
\newblock In {\em Proceedings of the 10th International Conference on Logic
  Programming}, pages 135--149, Budapest, Hungary, 1993. The MIT Press.

\bibitem{Bonwick-94}
J.~Bonwick.
\newblock {The Slab Allocator: An Object-Caching Kernel Memory Allocator}.
\newblock In {\em Proceedings of the Usenix Summer 1994 Technical Conference},
  pages 87--98, Boston, USA, 1994. Usenix Association.

\bibitem{Chen-95}
W.~Chen, T.~Swift, and D.~S. Warren.
\newblock {Efficient Top-Down Computation of Queries under the Well-Founded
  Semantics}.
\newblock {\em Journal of Logic Programming}, 24(3):161--199, 1995.

\bibitem{Chen-96}
W.~Chen and D.~S. Warren.
\newblock {Tabled Evaluation with Delaying for General Logic Programs}.
\newblock {\em Journal of the ACM}, 43(1):20--74, 1996.

\bibitem{Ciepielewski-91}
A.~Ciepielewski.
\newblock {Scheduling in Or-parallel Prolog Systems: Survey and Open Problems}.
\newblock {\em International Journal of Parallel Programming}, 20(6):421--451,
  1991.

\bibitem{Damasio-00}
C.~Damásio.
\newblock {A distributed tabling system}.
\newblock In {\em Proceedings of the 2nd Conference on Tabulation in Parsing
  and Deduction}, pages 65--75, Vigo, Spain, 2000.

\bibitem{Demoen-98}
B.~Demoen and K.~Sagonas.
\newblock {CAT: the Copying Approach to Tabling}.
\newblock In {\em Proceedings of the 10th International Symposium on
  Programming Language Implementation and Logic Programming}, number 1490 in
  Lecture Notes in Computer Science, pages 21--35, Pisa, Italy, 1998.
  Springer-Verlag.

\bibitem{Demoen-00}
B.~Demoen and K.~Sagonas.
\newblock {CHAT: The Copy-Hybrid Approach to Tabling}.
\newblock {\em Future Generation Computer Systems}, 16(7):809--830, 2000.

\bibitem{Freire-95}
J.~Freire, R.~Hu, T.~Swift, and D.~S. Warren.
\newblock {Exploiting Parallelism in Tabled Evaluations}.
\newblock In {\em Proceedings of the 7th International Symposium on Programming
  Languages: Implementations, Logics and Programs}, number 982 in Lecture Notes
  in Computer Science, pages 115--132, Utrecht, The Netherlands, 1995.
  Springer-Verlag.

\bibitem{Freire-96}
J.~Freire, T.~Swift, and D.~S. Warren.
\newblock {Beyond Depth-First: Improving Tabled Logic Programs through
  Alternative Scheduling Strategies}.
\newblock In {\em Proceedings of the Eight International Symposium on
  Programming Language Implementation and Logic Programming}, number 1140 in
  Lecture Notes in Computer Science, pages 243--258, Aachen, Germany, 1996.
  Springer-Verlag.

\bibitem{Guo-01}
Hai-Feng Guo and G.~Gupta.
\newblock {A Simple Scheme for Implementing Tabled Logic Programming Systems
  Based on Dynamic Reordering of Alternatives}.
\newblock In {\em Proceedings of the 17th International Conference on Logic
  Programming}, number 2237 in Lecture Notes in Computer Science, pages
  181--196, Paphos, Cyprus, 2001. Springer-Verlag.

\bibitem{Guo-02}
Hai-Feng Guo and G.~Gupta.
\newblock {Cuts in Tabled Logic Programming}.
\newblock In {\em Proceedings of the Colloquium on Implementation of Constraint
  and LOgic Programming Systems}, Copenhagen, Denmark, 2002.

\bibitem{Gupta-01}
G.~Gupta, E.~Pontelli, K.~Ali, M.~Carlsson, and M.~V. Hermenegildo.
\newblock {Parallel Execution of Prolog Programs: A Survey}.
\newblock {\em ACM Transactions on Programming Languages and Systems},
  23(4):472--602, 2001.

\bibitem{Hausman-PhD}
B.~Hausman.
\newblock {\em {Pruning and Speculative Work in OR-Parallel PROLOG}}.
\newblock PhD thesis, The Royal Institute of Technology, Stockholm, Sweden,
  1990.

\bibitem{Hermenegildo-Phd}
M.~V. Hermenegildo.
\newblock {\em {An Abstract Machine Based Execution Model for Computer
  Architecture Design and Efficient Implementation of Logic Programs in
  Parallel}}.
\newblock PhD thesis, University of Texas, Austin, Texas, USA, 1986.

\bibitem{Hermenegildo-91}
M.~V. Hermenegildo and K.~Greene.
\newblock {The \&-Prolog System: Exploiting Independent And-Parallelism}.
\newblock {\em New Generation Computing}, 9(3,4):233--257, 1991.

\bibitem{Hu-PhD}
R.~Hu.
\newblock {\em {Efficient Tabled Evaluation of Normal Logic Programs in a
  Distributed Environment}}.
\newblock PhD thesis, Department of Computer Science, State University of New
  York, Stony Brook, USA, 1997.

\bibitem{Johnson-99}
E.~Johnson, C.~R. Ramakrishnan, I.~V. Ramakrishnan, and P.~Rao.
\newblock {A Space Efficient Engine for Subsumption-Based Tabled Evaluation of
  Logic Programs}.
\newblock In {\em Proceedings of the 4th Fuji International Symposium on
  Functional and Logic Programming}, number 1722 in Lecture Notes in Computer
  Science, pages 284--300, Tsukuba, Japan, 1999. Springer-Verlag.

\bibitem{Kowalski-79}
R.~Kowalski.
\newblock {\em Logic for Problem Solving}.
\newblock Artificial Intelligence Series. North-Holland, 1979.

\bibitem{Kozen-83}
D.~Kozen.
\newblock {Results on the propositional $\mu$-calculus}.
\newblock {\em Theoretical Computer Science}, 27:333--354, 1983.

\bibitem{Castro-03}
D.~S.~Warren L.~F.~Castro.
\newblock {Approximate Pruning in Tabled Logic Programming}.
\newblock In {\em Proceedings of the 12th European Symposium on Programming},
  volume 2618 of {\em Lecture Notes in Computer Science}, pages 69--83, Warsaw,
  Poland, 2003. Springer Verlag.

\bibitem{Aurora-88}
E.~Lusk, R.~Butler, T.~Disz, R.~Olson, R.~Overbeek, R.~Stevens, D.~H.~D.
  Warren, A.~Calderwood, P.~Szeredi, S.~Haridi, P.~Brand, M.~Carlsson,
  A.~Ciepielewski, and B.~Hausman.
\newblock {The Aurora Or-Parallel Prolog System}.
\newblock In {\em Proceedings of the International Conference on Fifth
  Generation Computer Systems}, pages 819--830, Tokyo, Japan, 1988. Institute
  for New Generation Computer Technology.

\bibitem{Marques-00}
R.~Marques, T.~Swift, and J.~Cunha.
\newblock {An Architecture for a Multi-threaded Tabling Engine}.
\newblock In {\em Proceedings of the 2nd Conference on Tabulation in Parsing
  and Deduction}, pages 141--154, Vigo, Spain, 2000.

\bibitem{Mattern-89}
F.~Mattern.
\newblock {Global Quiescence Detection based on Credit Distribution and
  Recovery}.
\newblock {\em Information Processing Letters}, 30(4):195--200, 1989.

\bibitem{Michie-68}
D.~Michie.
\newblock {Memo Functions and Machine Learning}.
\newblock {\em Nature}, 218:19--22, 1968.

\bibitem{Milner-89}
R.~Milner.
\newblock {\em {Communication and Concurrency}}.
\newblock International Series in Computer Science. Prentice Hall, 1989.

\bibitem{Pontelli-97}
E.~Pontelli and G.~Gupta.
\newblock {Implementation Mechanisms for Dependent And-Parallelism}.
\newblock In {\em Proceedings of the 14th International Conference on Logic
  Programming}, pages 123--137, Leuven, Belgium, 1997. The MIT Press.

\bibitem{Ramakrishnan-00}
C.~R. Ramakrishnan, I.~V. Ramakrishnan, S.~Smolka, Y.~Dong, X.~Du,
  A.~Roychoudhury, and V.~Venkatakrishnan.
\newblock {XMC: A Logic-Programming-Based Verification Toolset}.
\newblock In {\em Proceedings of the 12th International Conference on Computer
  Aided Verification}, number 1855 in Lecture Notes in Computer Science, pages
  576--580, Chicago, Illinois, USA, 2000. Springer-Verlag.

\bibitem{Ramakrishnan-99}
I.~V. Ramakrishnan, P.~Rao, K.~Sagonas, T.~Swift, and D.~S. Warren.
\newblock {Efficient Access Mechanisms for Tabled Logic Programs}.
\newblock {\em Journal of Logic Programming}, 38(1):31--54, 1999.

\bibitem{Rao-96}
P.~Rao, C.~R. Ramakrishnan, and I.~V. Ramakrishnan.
\newblock {A Thread in Time Saves Tabling Time}.
\newblock In {\em Proceedings of the Joint International Conference and
  Symposium on Logic Programming}, pages 112--126, Bonn, Germany, 1996. MIT
  Press.

\bibitem{Rao-97}
P.~Rao, K.~Sagonas, T.~Swift, D.~S. Warren, and J.~Freire.
\newblock {XSB: A System for Efficiently Computing Well-Founded Semantics}.
\newblock In {\em Proceedings of the Fourth International Conference on Logic
  Programming and Non-Monotonic Reasoning}, number 1265 in Lecture Notes in
  Computer Science, pages 431--441, Dagstuhl, Germany, 1997. Springer-Verlag.

\bibitem{Rocha-PhD}
R.~Rocha.
\newblock {\em {On Applying Or-Parallelism and Tabling to Logic Programs}}.
\newblock PhD thesis, Computer Science Department, University of Porto, 2001.

\bibitem{Rocha-99a}
R.~Rocha, F.~Silva, and V.~{Santos Costa}.
\newblock {Or-Parallelism within Tabling}.
\newblock In {\em Proceedings of the First International Workshop on Practical
  Aspects of Declarative Languages}, number 1551 in Lecture Notes in Computer
  Science, pages 137--151, San Antonio, Texas, USA, 1999. Springer-Verlag.

\bibitem{Rocha-99b}
R.~Rocha, F.~Silva, and V.~{Santos Costa}.
\newblock {YapOr: an Or-Parallel Prolog System Based on Environment Copying}.
\newblock In {\em Proceedings of the 9th Portuguese Conference on Artificial
  Intelligence}, number 1695 in Lecture Notes in Artificial Intelligence, pages
  178--192, Évora, Portugal, 1999. Springer-Verlag.

\bibitem{Rocha-00}
R.~Rocha, F.~Silva, and V.~{Santos Costa}.
\newblock {YapTab: A Tabling Engine Designed to Support Parallelism}.
\newblock In {\em Proceedings of the 2nd Conference on Tabulation in Parsing
  and Deduction}, pages 77--87, Vigo, Spain, 2000.

\bibitem{Rocha-01}
R.~Rocha, F.~Silva, and V.~{Santos Costa}.
\newblock {On a Tabling Engine that Can Exploit Or-Parallelism}.
\newblock In {\em Proceedings of the 17th International Conference on Logic
  Programming}, number 2237 in Lecture Notes in Computer Science, pages 43--58,
  Paphos, Cyprus, 2001. Springer-Verlag.

\bibitem{Sagonas-PhD}
K.~Sagonas.
\newblock {\em {The SLG-WAM: A Search-Efficient Engine for Well-Founded
  Evaluation of Normal Logic Programs}}.
\newblock PhD thesis, Department of Computer Science, State University of New
  York, Stony Brook, USA, 1996.

\bibitem{Sagonas-98}
K.~Sagonas and T.~Swift.
\newblock {An Abstract Machine for Tabled Execution of Fixed-Order Stratified
  Logic Programs}.
\newblock {\em ACM Transactions on Programming Languages and Systems},
  20(3):586--634, 1998.

\bibitem{Sagonas-94}
K.~Sagonas, T.~Swift, and D.~S. Warren.
\newblock {XSB as an Efficient Deductive Database Engine}.
\newblock In {\em Proceedings of the ACM SIGMOD International Conference on the
  Management of Data}, pages 442--453, Minneapolis, Minnesota, USA, 1994. ACM
  Press.

\bibitem{Sagonas-96}
K.~Sagonas, T.~Swift, and D.~S. Warren.
\newblock {An Abstract Machine for Computing the Well-Founded Semantics}.
\newblock In {\em Proceedings of the Joint International Conference and
  Symposium on Logic Programming}, pages 274--288, Bonn, Germany, 1996. The MIT
  Press.

\bibitem{Costa-99b}
V.~{Santos Costa}.
\newblock {Optimising Bytecode Emulation for Prolog}.
\newblock In {\em Proceedings of Principles and Practice of Declarative
  Programming}, number 1702 in Lecture Notes in Computer Science, pages
  261--267, Paris, France, 1999. Springer-Verlag.

\bibitem{Costa-91}
V.~{Santos Costa}, D.~H.~D. Warren, and R.~Yang.
\newblock {Andorra-I: A Parallel Prolog System that Transparently Exploits both
  And- and Or-Parallelism}.
\newblock In {\em Proceedings of the 3rd ACM SIGPLAN Symposium on Principles
  and Practice of Parallel Programming}, pages 83--93, Williamsburg, Virginia,
  USA, 1991. ACM Press.

\bibitem{Shen-92}
K.~Shen.
\newblock {Exploiting Dependent And-parallelism in Prolog: The Dynamic
  Dependent And-Parallel Scheme (DDAS)}.
\newblock In {\em Proceedings of the Joint International Conference and
  Symposium on Logic Programming}, pages 717--731, Washington, DC, USA, 1992.
  MIT Press.

\bibitem{Zhou-01a}
Yi-Dong Shen, Li-Yan Yuan, Jia-Huai You, and Neng-Fa Zhou.
\newblock {Linear Tabulated Resolution Based on Prolog Control Strategy}.
\newblock {\em Theory and Practice of Logic Programming}, 1(1):71--103, 2001.

\bibitem{Swift-94b}
T.~Swift and D.~S. Warren.
\newblock {An abstract machine for SLG resolution: Definite Programs}.
\newblock In {\em Proceedings of the International Logic Programming
  Symposium}, pages 633--652, Ithaca, New York, 1994. The MIT Press.

\bibitem{Tamaki-86}
H.~Tamaki and T.~Sato.
\newblock {OLDT Resolution with Tabulation}.
\newblock In {\em Proceedings of the 3rd International Conference on Logic
  Programming}, number 225 in Lecture Notes in Computer Science, pages 84--98,
  London, 1986. Springer-Verlag.

\bibitem{Tarjan-72}
R.~E. Tarjan.
\newblock {Depth-First Search and Linear Graph Algorithms}.
\newblock {\em SIAM Journal on Computing}, 1(2):146--160, 1972.

\bibitem{xmc}
{The~XSB~Group}.
\newblock {LMC: The Logic-Based Model Checking Project}, 2003.
\newblock Available from \verb+http://www.cs.sunysb.edu/~lmc+.

\bibitem{xsb}
{The~XSB~Group}.
\newblock {The XSB Logic Programming System}, 2003.
\newblock Available from \verb+http://xsb.sourceforge.net+.

\bibitem{Tick-91}
E.~Tick.
\newblock {\em {Parallel Logic Programming}}.
\newblock The MIT Press, 1991.

\bibitem{Vieille-89}
L.~Vieille.
\newblock {Recursive Query Processing: The Power of Logic}.
\newblock {\em Theoretical Computer Science}, 69(1):1--53, 1989.

\bibitem{Warren-83}
D.~H.~D. Warren.
\newblock {An Abstract Prolog Instruction Set}.
\newblock Technical Note 309, SRI International, 1983.

\bibitem{Zhou-00}
Neng-Fa Zhou, Yi-Dong Shen, Li-Yan Yuan, and Jia-Huai You.
\newblock {Implementation of a Linear Tabling Mechanism}.
\newblock In {\em Proceedings of Practical Aspects of Declarative Languages},
  number 1753 in Lecture Notes in Computer Science, pages 109--123, Boston, MA,
  USA, 2000. Springer-Verlag.

\end{thebibliography}

\end{document}